\documentclass[10pt,twocolumn]{article}
\usepackage{calc}
\usepackage{amsmath,amssymb,amsthm,amscd}
\numberwithin{equation}{section}
\usepackage[bf]{caption}
\usepackage{longtable}
\usepackage{array}
\usepackage{enumerate}
\usepackage{float}
\usepackage{subfig}
\usepackage{multicol}
\usepackage{epsfig}
\usepackage{graphicx}
\usepackage{color}
\usepackage{cite}
\usepackage[width=1.09\textwidth]{caption}
\usepackage{tikz}
\usepackage{hyperref}
\usepackage{tikz,textcomp}
\usetikzlibrary{shapes,arrows,positioning,chains,calc}

\usepackage[utf8]{inputenc}
\usepackage[T1]{fontenc}
\usepackage[english]{babel}
\usepackage{xspace}
\usepackage{pifont}

\newcommand{\one}[0]{\ensuremath{\mathbf{1} }\xspace}
\newcommand{\two}[0]{\ensuremath{\mathbf{2} }\xspace}

\newcommand{\three}[0]{\ensuremath{\mathbf{3} }\xspace}

\newcommand{\beq}{\begin{equation}}
\newcommand{\eeq}{\end{equation}}
\newcommand{\bea}{\begin{eqnarray}}
\newcommand{\eea}{\end{eqnarray}}

\newtheorem{conj}{Conjecture}

\begin{document}
\twocolumn[{
\baselineskip=15pt
\begin{center}

\vspace{ 0.0cm}
{\Large {\bf Mordell-Weil Torsion in the Mirror of Multi-Sections}
}
\\[0pt]
\vspace{-0.1cm}
\bigskip
{{\bf Paul-Konstantin Oehlmann}$^{\,\text{a,b}}$,
{{\bf Jonas~Reuter}$^{\, \text{b}}$}, and \\
{{\bf Thorsten Schimannek}$^{\,\text{b}}$}
\bigskip }\\[3pt]
\vspace{0.cm}
{\it \small
${}^{\text{a}}$Physics Department, Robeson Hall, Virginia Tech, Blacksburg,~VA~24061,~USA\\[0.5cm]}
{\it \small
${}^{\text{b}}$ Bethe~Center~for~Theoretical~Physics, Physikalisches~Institut~der~Universit\"at~Bonn,\\ Nussallee~12,~53115~Bonn,~Germany\\[0.4cm]}

{\small oehlmann@th.physik.uni-bonn.de, jreuter@th.physik.uni-bonn.de, schimann@th.physik.uni-bonn.de}
\\[0.4cm]
\end{center}

\begin{center} {\bf Abstract } \end{center}
{
\small
We give further evidence that genus-one fibers with multi-sections are mirror dual to fibers with Mordell-Weil torsion.
In the physics of F-theory compactifications this implies a relation between models with a non-simply connected gauge group and those with discrete symmetries.
We provide a combinatorial explanation of this phenomenon for toric hypersurfaces.
In particular this leads to a criterion to deduce Mordell-Weil torsion directly from the polytope.
For all 3134 complete intersection genus-one curves in three-dimensional toric ambient spaces we confirm the conjecture by explicit calculation.
We comment on several new features of these models: The Weierstrass forms of many models can be identified by relabeling the coefficient sections.
This reduces the number of models to 1024 inequivalent ones.
We give an example of a fiber which contains only non-toric sections one of which becomes toric when the fiber is realized in a different ambient space.
Similarly a singularity in codimension one can have a toric resolution in one representation while it is non-toric in another.
Finally we give a list of 24 inequivalent genus-one fibers that simultaneously exhibit multi-sections and Mordell-Weil torsion in the Jacobian.
We discuss a self-mirror example from this list in detail.
}
\begin{flushright}
\parbox[t]{1.2in}{\textit{April, 2016}}
\end{flushright}
\vspace{.7cm}

}]
\section{Introduction}
\label{sec:introduction}
F-theory encodes the varying axio-dilaton profile of a type IIB string compactification in an elliptic or genus-one\footnote{In the following we will use the term genus-one fibration and also imply elliptic fibrations when the presence of a rational section is not important.}
fibration $Y$ over the physical base $B$ \cite{Vafa:1996xn, Morrison:1996pp}
\begin{align}
\pi: Y\rightarrow B \, .
\end{align}
In general the fiber degenerates over codimension one loci in the base which in the type IIB picture is interpreted as a stack of (p, q) 7-branes wrapping the divisor.
The divisors resolving the singularities correspond to the Cartan elements of a non-Abelian gauge algebra localized on the branes.
On the other hand Abelian gauge symmetries, the fundamental group of the gauge group and discrete symmetries are global properties of the fibration:
They are encoded in the non-torsional sections\cite{Morrison:1996pp,Morrison:2012ei}, torsional sections \cite{Aspinwall:1998xj,Mayrhofer:2014opa} and multi-sections \cite{Braun:2014oya,Morrison:2014era,Mayrhofer:2014haa}.

If the fibration admits a section, one of them can be used to single out a zero point and therefore a group law on each fiber.
This section is accordingly called the \textit{zero-section}.
Moreover, the group law lifts to the sections which then generate the \textit{Mordell-Weil group}.
In particular the rest of the sections will be either torsional or non-torsional under the group law.

The non-torsional sections generate the free part of the Mordell-Weil group.
They are mapped under a group homomorphism called the {\it Shioda map} \cite{Morrison:2012ei} into the group of vertical divisors in the fourfold.\footnote{More correctly the codomain of the Shioda-map is the N\'{e}ron-Severi group which is in the cases we are interested in equivalent to the Picard-group \cite{Mayrhofer:2014opa}.}
The Shioda map is up to base divisors of the form
\begin{align}
\sigma(S)=S-Z+ \sum_i a_i D_i\,,\,
\end{align}
where we take $Z$ as the zero-section.
Here $D_i$ are divisors on the ambient space that restrict on singular loci to the Cartan divisors of the gauge algebra $G$. The coefficients $a_i$ are rational numbers and related to the inverse Cartan matrix of $G$.

The expansion of the M-theory three-form $C_3$ along the Shioda map gives rise to the one-form potential of the $U(1)$ gauge symmetry in the effective field theory.
Moreover $\sigma$ can be used as a charge operator when intersected with matter curves that appear as fiber components resolving enhanced singularities over codimension two loci in the base. 

The torsional part of the Mordell-Weil group is generated by k-torsional sections and in analogy to the free part of the Mordell-Weil group
we can formulate a {\it torsion Shioda map} $\sigma^{(k)}_{t}$. \cite{Mayrhofer:2014opa}
In the fourfold geometry $\sigma^{(k)}_t(T)$ is up to base divisors a trivial $\mathbb{Q}$-divisor of the form
\begin{align}
\label{eqn:torsionShiodaMap}
\sigma^{(k)}_t(T)=T-Z+\frac{1}{k}\sum\limits_ic_i D_i \, ,
\end{align}
where $T$ is a $k$-torsional section when we take $Z$ as the zero-section.
The triviality reflects the fact that this is the image of a torsional section under a group homomorphism.
Since $\sigma^{(k)}_t(T)$ is a trivial $\mathbb{Q}$-divisor we find that intersections of curves with $\sum_ic_i D_i$ need to be multiples of $k$.
In particular this implies a condition on the weights of the matter curves and restricts the admissible representations of the gauge group $G$.
Globally this leads to a non-simply connected gauge group with fundamental group $\pi_1 (G) = \mathbb{Z}_k$ \cite{Aspinwall:1998xj}.\footnote{The presence of additional factors in $\pi_1(G)$ should correspond to the existence of multiple torsional sections that are linearly independent.}

On the other hand, by general arguments from quantum gravity, a discrete symmetry should always be understood as the discrete remnant of a $U(1)$, higgsed with a non-minimally charged multiplet \cite{Banks:2010zn}.
Geometrically a section that generated a part of the Mordell-Weil group combines with another $k$-section into a single $(k+1)$-section \cite{Morrison:2014era}.
The Shioda map becomes a generalized {\it discrete Shioda map} $\sigma^{(k+1)}_d$ that captures the $\mathbb{Z}_{k+1}$ quantum numbers of the matter curves \cite{Klevers:2014bqa,Grimm:2015wda}.

A fibration that only admits multi-sections has an associated \textit{Jacobian fibration} which leads to the same axio-dilaton profile but has a canonical zero section.
Moreover the genus-one fibration can be equipped with an action of the Jacobian fibration \cite{Cvetic:2015moa}.

The set of fibrations corresponding to the same Jacobian fibration and a specified action of the latter form a group that has been called the Tate-Shafarevich group in the F-Theory literature.\footnote{
See however footnote 15 in \cite{Bhardwaj:2015oru} for a clarification of the nomenclature.
}
In the following we will not refer to this mathematical formulation but always use the absence of a section and the presence of $k$-sections as our criterion for discrete symmetry.

All of these ingredients are necessary to construct realistic global F-theory models and understand the landscape of possible vacua \cite{Lin:2014qga,Mayrhofer:2014haa,Cvetic:2015txa}.
Although from the perspective of F-theory we view the genus-one curve as a fiber, 
that fiber is in particular a Calabi-Yau one-fold and has a mirror curve obtained by T-duality.
Surprisingly in \cite{Klevers:2014bqa} it was observed that when the fiber is realized as a hypersuface in a two-dimensional toric variety
an elliptic fiber with $\mathbb{Z}_k$ Mordell-Weil torsion is mirror-dual to a genus-one fiber with $k$-sections. 

Apart from the 16 hypersurface cases, one more example of this phenomenon has been given in \cite{Braun:2014qka} as a mirror-pair of complete intersections in three-dimensional toric ambient spaces.
This led to what we will refer to in the following as the \textit{mirror conjecture}:\footnote{This formulation contains some hindsight which will become clear below.}
\begin{conj}
Given a genus-one fiber $\mathcal{C}$ for which the Mordell-Weil group of the Jacobian contains torsion, the mirror dual is a genus-one fiber $\mathcal{C}'$ without a section and vice versa.
\end{conj}

If this conjecture were true it would be puzzling as mirror symmetry is only applied to the fiber but the exchanged properties are intrinsic to the full fibration.
\subsection*{Outline and summary}
In this paper we want to investigate the phenomenon of mirror symmetry in the fiber and its connection to 
torsion and multi-sections in more detail.

In Section \ref{sec:2DCase} we start by approaching the hypersurface case from the viewpoint of toric geometry.
Turning the argument of \cite{Mayrhofer:2014opa} upside down we take the existence of a divisor of the form \eqref{eqn:torsionShiodaMap}
as the criterion for Mordell-Weil torsion. This leads us to a combinatorial explanation of the mirror phenomenon in the case of hypersurfaces.

In Section \ref{sec:3DCase} we turn to genus-one curves realized as complete intersections in three-dimensional toric ambient spaces where however our combinatorial criterion
does not fully carry over. We then explicitly confirm the proposed duality for all of the 3134 nef partitions.
On the way, we find various relations among
different genus-one realizations and many new features: First, many of the 3134 nef partitions have equivalent Jacobians which can be 
reduced to exactly $1024$ inequivalent models by the algorithm presented in Appendix~\ref{app:equivalence}.

Moreover in Section \ref{sec:nontoric40} we discuss an example with only non-toric rational sections.
Here we construct the divisor class of the rational section from the divisor classes of two multi-sections.
The corresponding Weierstrass form can be obtained from an equivalent nef partition with a toric rational section.

Finally we give 24 examples of inequivalent genus-one fibers that simultaneously exhibit multi-sections and Mordell-Weil torsion in the Jacobian.
In Section \ref{sec:1220} we discuss one of those models in full detail and confirm the presence of a discrete symmetry and restricted representations by explicitly working out the matter content.
This is also the first example where a two-section wraps a full fiber component at codimension two.

Mordell-Weil torsion is dual to multi-sections in all of the cases and we find only two mirror pairs where
the degree of Mordell-Weil torsion does not match the degree of the multi-sections.


\section{Toric geometry and the mirror-conjecture for hypersurfaces}
\label{sec:2DCase}
In this section we show that the degree of the Mordell-Weil torsion and the presence of multi-sections can be read off directly from the
toric diagram and the polar polytope.\footnote{As we describe in Appendix \ref{app:nonAbelian}, also the non-Abelian gauge groups can be immediately deduced
from the polytopes.}
Moreover we show that the torsion Shioda map \eqref{eqn:torsionShiodaMap} follows, up to base divisors, immediately from this data.

\subsection{Toric geometry}
To fix conventions we will review some basic facts from toric geometry and the Batyrev mirror construction for hypersurfaces. However, we will not give a self-contained introduction and
state facts only in a generality that is necessary for our argument.
For more information we refer the reader to a vast amount of literature, e.g. \cite{Fulton:1993,Cox:2000vi,cox2011toric}.

Given a full-dimensional lattice polytope $\Delta\subset M_\mathbb{R}$ with $0\in\Delta$, where $M_\mathbb{R}$ is the real extension of a lattice $M\simeq\mathbb{Z}^d$, one can construct the \textit{polar polytope}
{
\small
\begin{align}
\Delta^\circ=\{v\in N_\mathbb{R}\,:\,\langle m,\,v\rangle\ge-1\text{ for all }m\in\Delta\cap M\}\,.
\end{align}
}
If $\Delta^\circ$ is also a lattice polytope then $\Delta$ and $\Delta^\circ$ are called reflexive and $(\Delta^\circ)^\circ=\Delta$.

From the polytope $\Delta^\circ$ one can build a fan $\Sigma$ corresponding to a fine star triangulation.
While in general many triangulations are possible the physics should be independent of this choice.
In particular in two dimensions the choice is unique.
We can then associate a toric variety $\mathbb{P}_{\Delta^\circ}$ to the fan $\Sigma$
as follows.
For each one-dimensional cone in $\Sigma$ there is a minimal generator\footnote{In a slight abuse of notation we will use $\rho_i$ to denote the minimal generator as well as the cone itself.} $\rho_i\in N,\,i=1,\dots,n$ to which one associates a homogeneous coordinate $z_i$ in $\mathbb{C}^n$.
Moreover, the generators $\rho_i$ satisfy $k$ linear relations $\sum_il^j_i\rho_i,\,j=1,\dots,k$.
Then we can construct
\begin{align}
\mathbb{P}_{\Delta^\circ}=\frac{\mathbb{C}^n\,\backslash\,S}{\left(\mathbb{C}^*\right)^k} \, ,
\end{align}
where the torus action $\left(\mathbb{C}^*\right)^k$ acts as
\begin{align}
z_i\rightarrow \lambda_j^{l^j_i}z_i,\,j=1,\dots,k,\,i=1,\dots,n\,.
\end{align}
Here $S$ is the union of subvarieties
\begin{align}
\bigcup\limits_\mathcal{I}\left(\bigcap\limits_{m\in\mathcal{I}}\{z_m=0\}\right)\subset\mathbb{C}^n\, ,
\end{align}
with the property that the $\{\rho_m\,:\,m\in\mathcal{I}\}$ do not form a cone in $\Sigma$.

A basis of Weil divisors compatible with the torus action is given by $D_i=\{z_i=0\},\,i=1,\dots,n$.
By our construction the variety is smooth and every Weil divisor is $\mathbb{Q}$-Cartier.
While the fan $\Sigma$ lives in $N_\mathbb{R}$ the lattice $M$ is still of importance.
If the line bundle associated to a Cartier divisor $D=\sum_ia_iD_i$ in $\mathbb{P}_{\Delta^\circ}$ is generated by its global sections, then the divisor can be entirely described in terms of a polytope $\Delta_D\in M_\mathbb{R}$ with
\begin{align}
\Delta_D=\{m\in M_\mathbb{R}\,:\,\langle m,\,\rho_i\rangle\ge-a_i\,\forall\,i=1,\dots,n\}\, .
\label{eqn:divtopoly}
\end{align}
A shift of the polytope by some lattice vector changes $D$ only by a trivial divisor and the scaled polytope $k\cdot\Delta_D$ corresponds to the divisor $k\cdot D$.

The points of $\Delta_D\cap M$ are in one-to-one correspondence with global sections of $\mathcal{O}_{\mathbb{P}_{\Delta^\circ}}(D)$ and a general section takes the form
\begin{align}
P_{\Delta_D}=\sum\limits_{m\in\Delta_D}\prod\limits_{i=1}^ns_m z_i^{\langle m,\,\rho_i\rangle+a_i}.
\label{eqn:polytoeqn}
\end{align}
with coefficients $s_m$.

In particular the polytope $\Delta$ corresponds to the anti-canonical divisor $-K=\sum_iD_i$ and the equation
\begin{align}
P_{\Delta}=\sum\limits_{m\in\Delta}\prod\limits_{i=1}^ns_mz_i^{\langle m,\,\rho_i\rangle+1}=0\, ,
\end{align}
cuts out a Calabi-Yau $(d-1)$-fold $X$.

The mirror Calabi-Yau $Y$ is obtained by exchanging the role of $\Delta$ and $\Delta^\circ$.
For our argument we need a second fact about the lattice $M$, namely that the points $m\in M$ are in one-to-one correspondence to trivial divisors
\begin{align}
\label{eq:trivialD}
D=\sum\limits_i\langle m,\,\rho_i\rangle D_i\sim 0 \, .
\end{align}

\subsection{Lattice refinement and Shioda maps}
In the following we use these tools to understand the mirror conjecture for hypersurfaces.
A k-torsional section becomes trivial upon multiplication with $k$ and as a group homomorphism the torsion Shioda map\footnote{In a moderate abuse of notation we will in the following use the name Shioda map for the image of a particular section. We will mostly neglect the contribution of base divisors.} must respect this property\cite{Mayrhofer:2014opa,Grimm:2015wda}, i.e.
\begin{align}
k\cdot\sigma_t^{(k)} \sim 0 \, .
\end{align}
Using Equation (\ref{eq:trivialD}) we see that $k\cdot\sigma_t^{(k)}$ corresponds to a dual lattice point $m \in M$ via
\begin{align}
k\cdot\sigma_t^{(k)}=\sum_i \langle m, \rho_i\rangle D_i \, .
\label{eq:trivial}
\end{align}

Turning the logic of \cite{Mayrhofer:2014opa} upside down, a trivial divisor restricts the admissible representations of the gauge group if and only if
the coefficients of divisors $D_i$ that intersect the curve have a non-trivial greatest common divisor $k$ that is not a divisor of the coefficients multiplying the resolution divisors.\footnote{This holds of course only up to linear equivalence.}

As was already noted in \cite{Braun:2013yti} the T-invariant divisors that intersect the curve correspond to the vertices of the polytope $\Delta^\circ$.
Thus we need $m\in M$ such that $\langle m,~\rho_i\rangle\in~k\mathbb{Z}$ for all vertices $\rho_i$ and $\langle m, \rho_l\rangle\notin k\mathbb{Z}$ for some $\rho_l$ that are not vertices.
In other words \textit{the vertices of the polytope span a lattice of index $k$ in $N$.}

We now want to show that this condition is indeed equivalent to the absence of a section in the dual geometry.

First note that if and only if the vertices span a lattice of index $k$ in $N$, then $\mathbb{P}_{\Delta^\circ}$ is covered by resolved $\mathbb{C}^2/\mathbb{Z}_k$ orbifold patches, i.e. the number of points in any edge of $\Delta^\circ$ is $l\cdot k+1$ for some $l\in\mathbb{N}$.

From the perspective of the mirror dual fiber $Y$ the polytope $\Delta^\circ$ describes the anticanonical divisor on the ambient space $-K_{\mathbb{P}_{\Delta}}=D_{\Delta^\circ}$.
In two dimensions divisors are curves and by adjunction the curve class of $Y$ is exactly $D_{\Delta^\circ}$.
Given any vertex $\rho\in\Delta^\circ$ we can shift the polytope $\Delta^\circ-\rho$ so that the origin becomes one of the new vertices.

Since every edge of $\Delta^\circ-\rho$ still contains a multiple of $k+1$ points it is proportional to a smaller lattice polytope $\Delta'$, i.e. $\Delta^\circ-\rho=k\cdot\Delta'$.
In terms of the associated divisors
\begin{align}
D_{\Delta^\circ}\sim D_{\Delta^\circ-\rho}=k\cdot D_{\Delta'} \, .
\end{align}

The intersection of every divisor with $D_{\Delta'}$ is integral and therefore the fiber has no section but only $k$-sections.
Using the toric cone theorem \cite{cox2011toric} the whole argument can be reversed to show that from general principles Mordell-Weil torsion is indeed dual to multi-sections for fibers constructed as hypersurfaces in toric varieties.

Although this finishes the proof, we want to note that the intersections of toric divisors in $\mathbb{P}_{\Delta}$ with the curve can be directly related to the number of points on the facets of $\Delta^\circ$.

Given a toric divisor associated to a generator $\rho_1$ we can choose one of the neighboring generators $\rho_2$.
Then we can calculate the intersection of $D_1$ with the curve in the patch with variables $z_1,\,z_2$ and all other variables set to one.
The only monomials that do not vanish for $z_1=0$ correspond to points on the facet dual to $\rho_1$. Moreover, since the facet is orthogonal to $\rho_1$ it cannot be orthogonal to $\rho_2$.
Therefore imposing $z_1=0$ the equation describing the hypersurface in the chosen patch becomes a polynomial in $z_2$ and every point on the facet dual to $\rho_1$ corresponds to a different exponent.
It is easy to see that it is in fact a polynomial of degree equal to the number of points on the dual facet minus one.
The latter is then equal to the number of intersections of $D_1$ with the curve.
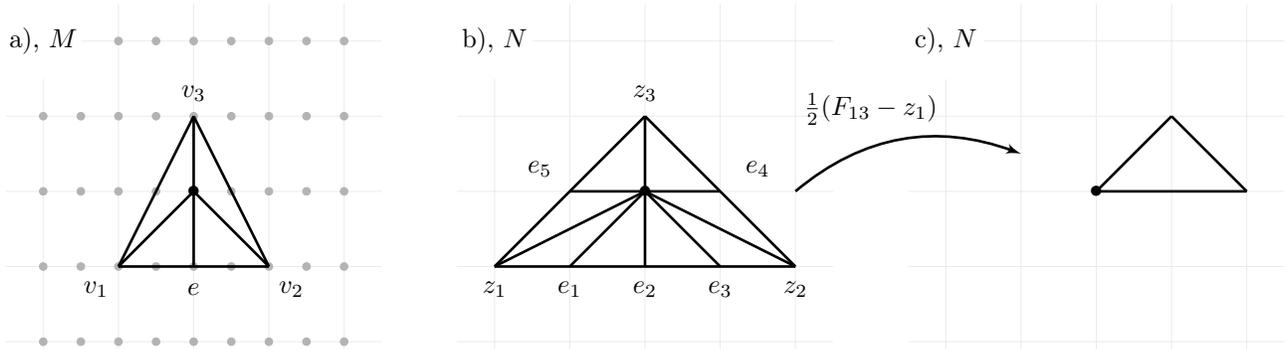
\begin{figure*}
\centering
\vspace{3cm}
\begin{tikzpicture}[remember picture,overlay,node distance=4mm, >=latex',block/.style = {draw, rectangle, minimum height=65mm, minimum width=83mm,align=center},]
\draw[help lines, overlay, lightgray!30] (-8.5,-0.5) grid (-3.5,4.5);
\draw[help lines, overlay, lightgray!30] (-2.5,-0.5) grid (2.5,4.5);
\draw[help lines, overlay, lightgray!30] (3.5,-0.5) grid (8.5,4.5);
\foreach \Point in {
(-7.5,0),(-6.5,0),(-5.5,0),(-4.5,0),
(-7.5,1),(-6.5,1),(-5.5,1),(-4.5,1),
(-7.5,2),(-6.5,2),(-5.5,2),(-4.5,2),
(-7.5,3),(-6.5,3),(-5.5,3),(-4.5,3),
(-6.5,4),(-5.5,4),(-4.5,4)
}{
    \draw[fill = white!30,black!30] \Point circle  [radius=0.05];
}
\foreach \Point in {
(-8,0),(-7,0),(-6,0),(-5,0),(-4,0),
(-8,1),(-7,1),(-6,1),(-5,1),(-4,1),
(-8,2),(-7,2),(-6,2),(-5,2),(-4,2),
(-8,3),(-7,3),(-6,3),(-5,3),(-4,3),
(-7,4),(-6,4),(-5,4),(-4,4)
}{
    \draw[fill = black!30,black!30] \Point circle  [radius=0.05];
}
\draw [line width=1] (-7,1) -- (-6,3);
\draw [line width=1] (-5,1) -- (-6,3);
\draw [line width=1] (-5,1) -- (-7,1);
\draw [line width=1] (-7,1) -- (-6,2);
\draw [line width=1] (-6,1) -- (-6,2);
\draw [line width=1] (-5,1) -- (-6,2);
\draw [line width=1] (-6,2) -- (-6,3);
\draw [line width=1] (-2,1) -- (0,3);
\draw [line width=1] (-2,1) -- (2,1);
\draw [line width=1] (-2,1) -- (0,2);
\draw [line width=1] (-1,1) -- (0,2);
\draw [line width=1] (0,1) -- (0,2);
\draw [line width=1] (1,1) -- (0,2);
\draw [line width=1] (2,1) -- (0,2);
\draw [line width=1] (2,1) -- (0,3);
\draw [line width=1] (-1,2) -- (1,2);
\draw [line width=1] (0,2) -- (0,3);
\draw [line width=1] (6,2) -- (8,2);
\draw [line width=1] (6,2) -- (7,3);
\draw [line width=1] (7,3) -- (8,2);

\foreach \Point in {(-6,2), (0,2), (6,2)}{
    \node at \Point {\textbullet};
}
\draw[->, black, overlay, line width=1] (2,2) to [out=40,in=160] (5,2.5);
\node[overlay] at (3, 3.1) {$\frac12(F_{13}-z_1)$};

\draw [fill=white,white] (-8.5,3.5) rectangle (-7.5,4.5);
\draw [fill=white,white] (-2.5,3.5) rectangle (-1.5,4.5);
\draw [fill=white,white] (3.5,3.5) rectangle (4.5,4.5);

\node[overlay] at (-7.3, .7) {$v_1$};
\node[overlay] at (-4.7, .7) {$v_2$};
\node[overlay] at (-6, 3.3) {$v_3$};
\node[overlay] at (-6, .7) {$e$};

\node[overlay] at (-8,4) {a), $M$};
\node[overlay] at (-2,4) {b), $N$};
\node[overlay] at (4,4) {c), $N$};

\node[overlay] at (-1.4,2.3) {$e_5$};
\node[overlay] at (1.5,2.3) {$e_4$};
\node[overlay] at (0,3.3) {$z_3$};
\node[overlay] at (-2,0.7) {$z_1$};
\node[overlay] at (2,0.7) {$z_2$};
\node[overlay] at (-1,0.7) {$e_1$};
\node[overlay] at (0,0.7) {$e_2$};
\node[overlay] at (1,0.7) {$e_3$};

\end{tikzpicture}
\caption{
The dual pair of polytopes $F_4$ and $F_{13}$ is shown in a) and b).
We also indicated the toric fan obtained from a complete star triangulation and labelled the homogeneous variables.
The gray dots in a) mark the lattice dual to the vertices of $F_{13}$.
A shift followed by a rescaling of $F_{13}$ leads to polytope $\Delta'=\frac12\left(F_{13}-z_1\right)$ shown in c).
}
\label{fig:f4f13}
\end{figure*}
\subsection*{Example: $F_{4}$ and $F_{13}$}
We illustrate the above with the dual pair of polytopes $F_{4}$ and $F_{13}$ shown in Figure~\ref{fig:f4f13}, a) and b).\footnote{For the numbering of two-dimensional reflexive polytopes we follow the conventions from \cite{Klevers:2014bqa} although
this deviates from the ids in the PALP database\cite{PALP}.
}
First look at the toric variety $\mathbb{P}_{F_{13}}$.
The vertices of $F_{13}$ are given by
\begin{align}
z_1=(-2,\,-1)\, ,\,z_2=(2,\,-1)\, ,\,z_3=(0,\,1)\, ,
\end{align}
and there are five points $e_i,\,i=1,...,5$ in the relative interior of facets.
While $[z_3]$ leads to a two-section $[z_1]$ and $[z_2]$ each intersect the curve once. The divisors $[e_i]$ do not intersect the curve and provide the Cartan elements of the $(SU(4)\times SU(2)^2)/\mathbb{Z}_2$ gauge group.
The sections $[z_1]$ and $[z_2]$ are not linearly dependent so one might expect a factor in the $U(1)$ gauge symmetry.
However, one easily sees that the vertices span a sublattice of index $2$ in $N$.
The dual lattice $\tilde{M}$ is generated by
\begin{align}
b_1=(1/2,\,0)\, ,\,b_2=(0,\,1) \, ,
\end{align}
and we conclude that taking one of the divisors
as the zero section the other becomes a torsional section.
In particular we can use the element $b_1\in \tilde{M}$ to immediately write down the torsion Shioda map 
\begin{align}
\sigma_t^{(2)}=[z_2]-[z_1]+\frac{1}{2}\left(-[e_1]+[e_3]+[e_4]-[e_5]\right) \, ,
\end{align}
up to base divisors via (\ref{eq:trivial}), in agreement with \cite{Mayrhofer:2014opa, Reuter:2015loa}.

That the vertices span a lattice of index $2$ implies that polytope $F_{13}$ can be shifted and rescaled
to the polytope
\begin{align}
\Delta'=\frac12\left(F_{13}-z_1\right) \, ,
\end{align}
shown in Figure~\ref{fig:f4f13}, c).
From the perspective of the ambient space $\mathbb{P}_{F_{4}}$ the polytope $F_{13}$ describes the anticanonical divisor
\begin{align}
-K=[v_1]+[v_2]+[v_3]+[e] \, .
\end{align}
Using the linear equivalences
\begin{align}
[v_1]\sim[v_2],\quad [v_3]\sim[v_1]+[v_2]+[e]\, ,
\end{align}
we get
\begin{align}
-K\sim4[v_1]+2[e]=2\cdot\left(2[v_1]+[e]\right) \, ,
\end{align}
where the divisor $D'=2[v_1]+[e]$ indeed corresponds to $\Delta'$ via (\ref{eqn:divtopoly}).
This is equivalent to the fact that the fiber in $\mathbb{P}_{F_{4}}$ is a genus-one curve admitting only two-sections.
As an interesting aside we note that for hypersurfaces the anticanonical divisor does also provide a charge assignment for the discrete symmetry.
It is naturally orthogonalized with respect to the resolution divisors and can be rescaled to match the discrete Shioda maps given in \cite{Klevers:2014bqa}.
\section{Evidence in CICYs}
\label{sec:3DCase}
The mirror construction for hypersurfaces was generalized by Batyrev and Borisov to complete intersections in toric ambient spaces \cite{borisov, Batyrev:1994pg}.
A complete intersection Calabi-Yau in $\mathbb{P}_{\Delta^\circ}$ is described in terms of a nef-partition
of the anti-canonical divisor.
Following \cite{Braun:2014qka} we define a nef partition as sets of polytopes $\Delta_1,...,\Delta_n$ and $\nabla_1,...,\nabla_n$ such that
\begin{equation}
  \begin{aligned}
&\Delta\phantom{^\circ}=\Delta_1+...+\Delta_n\, ,\quad&\Delta^\circ=\langle\nabla_1,...,\nabla_n\rangle \, ,\phantom{ll}\\
&\nabla^\circ=\langle\Delta_1,...,\Delta_n\rangle\, ,\quad&\nabla\phantom{^\circ}=\nabla_1+...+\nabla_n,
  \end{aligned}
\end{equation}
where $+$ denotes the Minkowski-sum and angular brackets the convex hull.
The intersection of the equations $P_{\Delta_i}=0$ associated to the polytopes $\Delta_i,\,i=1,...,n$ via (\ref{eqn:polytoeqn}) cuts out
a Calabi-Yau variety of codimension $n$ in $\mathbb{P}_{\Delta^\circ}$.
The mirror Calabi-Yau is cut out by $P_{\nabla_i}=0$ in $\mathbb{P}_{\nabla^\circ}$.

The combinatorial explanation for the relation between Mordell-Weil torsion and multi-sections
under mirror symmetry given above does rely on special features of two-dimensional toric varieties
and the Batyrev mirror construction for hypersurfaces.
For at least three reasons it does not readily generalize to complete intersections in higher dimensional toric varieties:
\begin{enumerate}
\item A vertex might correspond to a toric divisor that does not intersect the curve.
\item A singularity of the curve might not be torically resolved (see Appendix \ref{app:nonAbelian}).
\item Divisors are not curves.
\end{enumerate}
We can refine our torsion criterion to the following conjecture that we expect to hold in general:
\begin{conj}
A curve $\mathcal{C}$ constructed as a complete intersection in a toric ambient space such that all the codimension one loci are torically resolved does exhibit Mordell-Weil torsion of degree $k$ in the Jacobian if and only if 
the one dimensional generators of the fan $\{\rho_i\,:\,D_i\cdot C\ne 0\}$ span a sublattice $\tilde{M}\supset M$ of index $k$.
Up to base divisors a point $m\in\tilde{M}\backslash M$ corresponds to a torsion Shioda map $\sigma_t^{(k)}$ via
\begin{align}
\sigma_t^{(k)}=\sum\limits_{\rho_i\in\Sigma(1)}\langle m,\,\rho_i\rangle\cdot D_i\,.
\end{align}
\end{conj}
It is not clear to us how this can be used to generalize the argument given in Section \ref{sec:2DCase}.

One could expect that the phenomenon itself is therefore restricted to hypersurfaces.
However, a more general example was given in \cite{Braun:2014qka} where the authors studied complete intersections in three-dimensional toric ambient spaces.
They found that nef \href{http://wwwth.mpp.mpg.de/members/jkeitel/weierstrass/data/0_0.txt}{(0,~0)}\footnote{The numbering of polytopes and nef partitions
follows the PALP database and is explained in \cite{Braun:2014qka}. In the following we will often refer to the data of particular nefs which can be found on the website of Jan Keitel under
\url{http://wwwth.mpp.mpg.de/members/jkeitel/weierstrass/data/[polyId]_[nefId].txt}.
} is a genus-one fiber which only admits four-sections while the mirror dual nef \href{http://wwwth.mpp.mpg.de/members/jkeitel/weierstrass/data/3415_0.txt}{(3415,~0)}  exhibits $\mathbb{Z}_4$ Mordell-Weil torsion.

As a main result the authors calculated the Weierstrass coefficients of the Jacobians for all of the 3134 complete intersection fibers in three-dimensional toric ambient spaces.
They also determined the toric Mordell-Weil group and the vanishing loci of the discriminant.
To check if the duality holds for this more general class of fibers we calculated the intersection of all toric divisors with the curves.
Combining this with the information determined by \cite{Braun:2014qka} we find that
Mordell-Weil torsion $\mathbb{Z}_k$ does indeed imply $k$-sections in the mirror and vice versa in 3086 of the cases\footnote{
We provide the Sage \cite{Sage, sage_polytope, sage_toric} code to do the comparison and additional data on
\url{http://www.th.physik.uni-bonn.de/klemm/data.php}.
}.

The reasons why the conjecture seems to fail 46 times are twofold:
\begin{enumerate}
\item In nef \href{http://wwwth.mpp.mpg.de/members/jkeitel/weierstrass/data/4_0.txt}{(4,~0)} we only find toric two- and three-sections although the mirror nef \href{http://wwwth.mpp.mpg.de/members/jkeitel/weierstrass/data/3013_1.txt}{(3013,~1)} does not exhibit Mordell-Weil torsion.
We show below how one can show the existence of a non-toric section in \href{http://wwwth.mpp.mpg.de/members/jkeitel/weierstrass/data/4_0.txt}{(4,~0)}  which turns a superficial contradiction into additional evidence for the conjecture.
\item There are 45 fibers that only admit multi-sections while the mirror lacks a section as well.
We checked that in all of these cases there is a torsional section in the Weierstrass model by constructing a birational map into the generic forms listed in \cite{Aspinwall:1998xj}.
These are the first examples of genus-one fibers in F-theory that exhibit Mordell-Weil torsion in the Jacobian.
The models are shown in the following table:
\begin{center}
\renewcommand*{\arraystretch}{0.8}
{\small
\begin{tabular}{c|c c c|c}
\multicolumn{5}{c}{Dual pairs with two-torsion \textit{and} two-sections}\\
\multicolumn{5}{c}{}\\
\multicolumn{2}{c}{\small$\leftarrow$ mirror dual $\rightarrow$}&&\multicolumn{2}{c}{\small$\leftarrow$ mirror dual $\rightarrow$}\\
\multicolumn{5}{c}{}\\
$(152,\,0)$&$(195,\,4)$&&$(8,\,0)$&$(609,\,0)$\\&&&&\\[-1.6ex]
$(29,\,2)$&$(577,\,0)$&&$(34,\,0)$&$(321,\,1)$\\&&&&\\[-1.6ex]
$(39,\,0)$&$(335,\,0)$&&$(56,\,2)$&$(356,\,2)$\\&&&&\\[-1.6ex]
$(78,\,2)$&$(266,\,0)$&&$(108,\,0)$&$(161,\,1)$\\&&&&\\[-1.6ex]
$(129,\,0)$&$(129,\,1)$&&$(150,\,1)$&$(208,\,1)$\\\multicolumn{5}{c}{}\\[-1ex]
\multicolumn{2}{c}{$(152,\,1)$ self-mirror} & & \multicolumn{2}{c}{$(122,\,0)$ self-mirror}
\end{tabular}
}
\end{center}
We describe in the next section that there exist equivalent nefs for many of these models leading to the total number of 45.
As an example we provide a detailed analysis of nef \href{http://wwwth.mpp.mpg.de/members/jkeitel/weierstrass/data/122_0.txt}{(122,~0)} in Section \ref{sec:1220} and show that the gauge group is $\left(\text{SU}(2)^4/\mathbb{Z}_2\right)\times\mathbb{Z}_4$\footnote{
The discrete symmetry is enhanced by the center of the non-Abelian gauge group as we describe below.
}.
We give the discrete Shioda map and use Conjecture 2 to write down a generalized torsion Shioda map in terms of multi-sections.
\end{enumerate}
We want to stress that \textit{in all but two cases the degree of the Mordell-Weil torsion does match the degree of the multi-sections in the mirror}.
For two mirror pairs one of the fibers only admits four-sections while the Mordell-Weil torsion in the mirror is of degree $2$.
This contradicts a naive identification of the degrees and requires further investigation.
The two pairs are shown in the following table:
\begin{center}\renewcommand*{\arraystretch}{0.8}
{\small
\begin{tabular}{c|c}
\multicolumn{2}{c}{$\leftarrow$ mirror dual $\rightarrow$}\\
\multicolumn{2}{c}{}\\
nef $(5,\,3)$&nef $(2069,\,0)$ \\\hline\\[-1.8ex]
{\color{red}four-sections}&{\color{red}$\mathbb{Z}_2$ torsion}\\
no torsion&one-sections\\
&\\
nef $(21,\,1)$&nef $(488,\,0)$\\\hline\\[-1.8ex]
{\color{red}four-sections}&{\color{red}$\mathbb{Z}_2$ torsion}\\
$\mathbb{Z}_2$ torsion&two-sections
\end{tabular}
}
\end{center}
Although we do not know how to remedy this mismatch, we note that \href{http://wwwth.mpp.mpg.de/members/jkeitel/weierstrass/data/21_1.txt}{(21,~1)} higgses to \href{http://wwwth.mpp.mpg.de/members/jkeitel/weierstrass/data/5_3.txt}{(5,~3)}.
This suggests that there might be a common explanation for both fibers.
Moreover both geometries have non-split singularities in codimension one which we illustrate
in the example of \href{http://wwwth.mpp.mpg.de/members/jkeitel/weierstrass/data/5_3.txt}{(5,~3)} in Appendix \ref{app:nonAbelian}.

\subsection{Redundancy in CICYs}
\label{sec:redundancy}
Examining the resulting Weierstrass forms of the 3134 nef partitions shows a high redundancy.
Looking at the complete intersection equations of nef \href{http://wwwth.mpp.mpg.de/members/jkeitel/weierstrass/data/3013_0.txt}{(3013,~0)}
and \href{http://wwwth.mpp.mpg.de/members/jkeitel/weierstrass/data/3013_1.txt}{(3013,~1)} reveals that there exist nef partitions which are not
identical but lead to the same elliptic curves. 

Since the nef partitions are not equal the dual partitions correspond in general to
complete intersections in different ambient spaces.
In fact \href{http://wwwth.mpp.mpg.de/members/jkeitel/weierstrass/data/3013_0.txt}{(3013,~0)} is dual to \href{http://wwwth.mpp.mpg.de/members/jkeitel/weierstrass/data/4_0.txt}{(4,~0)} while the mirror of \href{http://wwwth.mpp.mpg.de/members/jkeitel/weierstrass/data/3013_1.txt}{(3013,~1)} is \href{http://wwwth.mpp.mpg.de/members/jkeitel/weierstrass/data/5_1.txt}{(5,~1)}.

Mirror symmetry identifies the complex structure and K\"ahler moduli space.
The complex structure moduli spaces of \href{http://wwwth.mpp.mpg.de/members/jkeitel/weierstrass/data/4_0.txt}{(4,~0)} and \href{http://wwwth.mpp.mpg.de/members/jkeitel/weierstrass/data/5_1.txt}{(5,~1)} are therefore both
identified with the K\"ahler moduli space of \href{http://wwwth.mpp.mpg.de/members/jkeitel/weierstrass/data/3013_0.txt}{(3013,~0)}$\sim$\href{http://wwwth.mpp.mpg.de/members/jkeitel/weierstrass/data/3013_1.txt}{(3013,~1)}.
We thus expect \href{http://wwwth.mpp.mpg.de/members/jkeitel/weierstrass/data/4_0.txt}{(4,~0)} and \href{http://wwwth.mpp.mpg.de/members/jkeitel/weierstrass/data/5_1.txt}{(5,~1)} to have the same Weierstrass form up to
permutations of the coefficient labels.

Since both families are parametrized by $18$ coefficients and the Weierstrass functions are huge it seems to be a daunting task to find the right permutation.
We therefore describe an algorithm to find the correct identification in Appendix \ref{app:equivalence}.

A more thorough analysis shows that the complete intersections fall into 1024 equivalence classes.\footnote{We provide the complete list of equivalence classes and additional data on \url{http://www.th.physik.uni-bonn.de/klemm/data.php}.}
This is particularly interesting because it can happen that a section that is toric in one representative might be non-toric in the other.
One example is nef \href{http://wwwth.mpp.mpg.de/members/jkeitel/weierstrass/data/4_0.txt}{(4,~0)}$\sim$\href{http://wwwth.mpp.mpg.de/members/jkeitel/weierstrass/data/5_1.txt}{(5,~1)} which we describe below.

Moreover, a model which has a codimension one singularity that is torically resolved can have another realization in which the resolution is non-toric.
This happens for \href{http://wwwth.mpp.mpg.de/members/jkeitel/weierstrass/data/108_0.txt}{(108,~0)}$\sim$\href{http://wwwth.mpp.mpg.de/members/jkeitel/weierstrass/data/149_2.txt}{(149,~2)}.
Both geometries describe genus-one fibers without a section and exhibit Mordell-Weil torsion.
Using the heuristics we describe in Appendix \ref{app:nonAbelian}, one can see that the $I_2$ singularity in \href{http://wwwth.mpp.mpg.de/members/jkeitel/weierstrass/data/108_0.txt}{(108,~0)} has a toric resolution while
the $I_2$ singularity in \href{http://wwwth.mpp.mpg.de/members/jkeitel/weierstrass/data/149_2.txt}{(149,~2)} does not.
In particular the criterion summarized in Conjecture 2 fails to see the Mordell-Weil torsion when applied to \href{http://wwwth.mpp.mpg.de/members/jkeitel/weierstrass/data/149_2.txt}{(149,~2)}.\footnote{
It would be interesting to follow the argument in \cite{Berglund:1998va} and interpret a non-toric resolution not as a deficit in the description but
as a physical freezing of the associated modulus by background flux.}

\subsection{Non-toric section in \href{http://wwwth.mpp.mpg.de/members/jkeitel/weierstrass/data/4_0.txt}{(4,~0)}}
\label{sec:nontoric40}
As we described above nef \href{http://wwwth.mpp.mpg.de/members/jkeitel/weierstrass/data/4_0.txt}{(4,~0)} does not have a toric section although there is no torsion in the Mordell-Weil group of its dual nef \href{http://wwwth.mpp.mpg.de/members/jkeitel/weierstrass/data/3013_1.txt}{(3013,~1)}.
We want to construct a non-toric section to show that this mirror pair does not contradict Conjecture 1.
The nef partition \href{http://wwwth.mpp.mpg.de/members/jkeitel/weierstrass/data/4_0.txt}{(4,~0)} is described by the set of polytopes
\begin{align}
\nabla_1=\langle z_0,\,z_3,\,z_4,\,z_6\rangle \, ,\quad\nabla_2=\langle z_1,\,z_2,\,z_5,\,z_8\rangle \, ,
\end{align}
with the points $z_i$ given in the following table:
\begin{center}
{\small
\begin{tabular}{ccccc}
$z_0$&$z_1$&$z_2$&$z_3$&$z_4$\\\hline
$\left( \begin{array}{c} 1 \\ 0 \\ 0 \end{array}\right)$&
$\left( \begin{array}{c} -1 \\ 0 \\ 0 \end{array}\right)$&
$\left( \begin{array}{c} 0 \\ 1 \\ 0 \end{array}\right)$&
$\left( \begin{array}{c} 0\\ 0 \\ 1 \end{array}\right)$&
$\left( \begin{array}{c} 0 \\ -1 \\ -1 \end{array}\right)$
\end{tabular}
}
\end{center}
One can see that the ambient space is $\mathbb{P}^1\times\mathbb{P}^2$.
In particular the toric divisors satisfy the linear equivalence relations
\begin{align}
[z_0]\sim[z_1]\,,\quad[z_2]\sim[z_3]\sim[z_4] \, .
\end{align}
The intersections with the curve $\mathcal{C}$ are therefore
determined by $\mathcal{C}\cdot[z_0]=2$ and $\mathcal{C}\cdot[z_2]=3$.

There is no toric section but along the lines of \cite{Braun:2013yti} one can argue that there exists a non-toric section $s$ that lies
in the homology class of the difference of a three- and a two-section. For the argument to work one needs to choose a base. As in \cite{Braun:2013yti}
we choose $\mathcal{B}=\mathbb{P}^2$ where the coordinates are in the divisor class $H_B$. 
Then the divisor $[s]=[z_2]-[z_0]+H_B$ intersects the fiber once and since the fibration does not have any singular loci in codimension one there are no further constraints on the intersections.

Using {\it cohomcalg} \cite{Blumenhagen:2010pv} we calculate the number of sections in this class and find $H^0(\mathcal{C}, \mathcal{O}([s]))=1$.
We have therefore shown the existence of a non-toric section.
This also implies that nef \href{http://wwwth.mpp.mpg.de/members/jkeitel/weierstrass/data/4_0.txt}{(4,~0)} is equivalent to nef \href{http://wwwth.mpp.mpg.de/members/jkeitel/weierstrass/data/5_1.txt}{(5,~1)} because both share the same Weierstrass equation.
Moreover, the section that is non-toric in \href{http://wwwth.mpp.mpg.de/members/jkeitel/weierstrass/data/4_0.txt}{(4,~0)} is torically realized in \href{http://wwwth.mpp.mpg.de/members/jkeitel/weierstrass/data/5_1.txt}{(5,~1)}.\footnote{
We find an additional non-toric rational section in \href{http://wwwth.mpp.mpg.de/members/jkeitel/weierstrass/data/5_1.txt}{(5,~1)} that can be constructed explicitly from a tangent to the curve as in \cite{Klevers:2014bqa}.
Together with the toric rational section this generates a Mordell-Weil group of rank one which leads to the gauge group $\mathrm{U}(1)$.
In fact \href{http://wwwth.mpp.mpg.de/members/jkeitel/weierstrass/data/5_1.txt}{(5,~1)} higgses to nef \href{http://wwwth.mpp.mpg.de/members/jkeitel/weierstrass/data/0_0.txt}{(0,~0)} so that we expect the presence of an abelian gauge group that is broken to the discrete $\mathrm{Z}_4$ symmetry in \href{http://wwwth.mpp.mpg.de/members/jkeitel/weierstrass/data/0_0.txt}{(0,~0)}.
On the other hand \href{http://wwwth.mpp.mpg.de/members/jkeitel/weierstrass/data/5_1.txt}{(5,~1)} can be obtained by higgsing two times from the $\mathrm{U}(1)^3$ theory described in \cite{Cvetic:2013qsa}.
}


\section{ The self-dual nef \href{http://wwwth.mpp.mpg.de/members/jkeitel/weierstrass/data/122_0.txt}{(122,~0)} }
\label{sec:1220}
In the final example we consider the genus-one curve given by the nef-partition \href{http://wwwth.mpp.mpg.de/members/jkeitel/weierstrass/data/122_0.txt}{(122,~0)}.
This is in an equivalence class with \href{http://wwwth.mpp.mpg.de/members/jkeitel/weierstrass/data/122_1.txt}{(122,~1)}, \href{http://wwwth.mpp.mpg.de/members/jkeitel/weierstrass/data/215_2.txt}{(215,~2)} and \href{http://wwwth.mpp.mpg.de/members/jkeitel/weierstrass/data/215_5.txt}{(215,~5)}.
The mirror dual nef is \href{http://wwwth.mpp.mpg.de/members/jkeitel/weierstrass/data/215_2.txt}{(215,~2)} and we see that the curve is therefore self-dual.\footnote{The Weierstrass coefficients of \href{http://wwwth.mpp.mpg.de/members/jkeitel/weierstrass/data/215_2.txt}{(215,~2)} can be mapped into those of \href{http://wwwth.mpp.mpg.de/members/jkeitel/weierstrass/data/122_0.txt}{(122,~0)} by substituting
{
\begin{align}
a_0&\rightarrow a_8\, ,\,a_1\rightarrow a_9\, ,\,a_2\rightarrow a_5\, ,\,a_3\rightarrow a_4\, ,\,a_4\rightarrow a_3\, ,\nonumber\\
a_5&\rightarrow a_9\, ,\,a_6\rightarrow a_6\, ,\,a_7\rightarrow a_2\, ,\,a_8\rightarrow a_1\, ,\,a_9\rightarrow a_0\, .\nonumber
\end{align}
}
}
The nef partition \href{http://wwwth.mpp.mpg.de/members/jkeitel/weierstrass/data/122_0.txt}{(122,~0)} is specified by
\begin{align}
\nabla_1=\langle z_0,\,z_3,\,z_4,\,z_6\rangle\, ,\quad\nabla_2=\langle z_1,\,z_2,\,z_5,\,z_8\rangle\, ,
\end{align}
with the points $z_i$ given in the following table:
\begin{center}
{\small
\begin{tabular}{cccc}
$z_0$&$z_1$&$z_2$&$z_3$\\\hline
$\left( \begin{array}{c} 1 \\ 0 \\ 0 \end{array}\right)$&
$\left( \begin{array}{c} 0 \\ 1 \\ 0 \end{array}\right)$&
$\left( \begin{array}{c} 0 \\ 0 \\ 1 \end{array}\right)$&
$\left( \begin{array}{c} -1\\ -1 \\ -1 \end{array}\right)$\\
&&&\\
$z_4$&$z_5$&$z_6$&$z_8$\\\hline
$\left( \begin{array}{c} 1 \\ -1 \\ -1 \end{array}\right)$&
$\left( \begin{array}{c} 0 \\ 2 \\ 1 \end{array}\right)$&
$\left( \begin{array}{c} 0 \\ -1 \\ -1 \end{array}\right)$&
$\left( \begin{array}{c} 0 \\ 1 \\ 1 \end{array}\right)$
\end{tabular}
}
\end{center}
The corresponding equations are
\begin{align}
\begin{split}
p_1 =\,&  a_2z_0^2 z_4^2 z_6 + a_4 z_0 z_1 z_5^2 z_8 + a_1 z_0 z_3 z_4 z_6\\
+\,&  a_3 z_0 z_2 z_5 z_8 + a_0 z_3^2 z_6 \, ,\, \\
p_2 =\,& a_9 z_0 z_1 z_4^2 z_6 + a_8 z_1^2 z_5^2 z_8 + a_6 z_1 z_3 z_4 z_6 \\ 
+\,& a_7 z_1 z_2 z_5 z_8 + a_5 z_2^2 z_8 \, .
\end{split}
\end{align}
There are two maximal star triangulations of this polytope and we choose the one leading to the Stanley-Reisner ideal
\begin{align}
\begin{split}
SR=\langle&z_3 z_4,\,z_0 z_3,\,z_1 z_8,\,z_1 z_2,\,z_4 z_5,\\
&z_2 z_5,\,z_5 z_6,\,z_4 z_8,\,z_6 z_8,\,z_0 z_6\rangle \, .
\end{split}
\end{align}
The toric divisors $[z_0],\,[z_1],\,[z_6],\,[z_8]$ do not intersect the curve while $[z_i],\,i\in\{2,\dots,5\}$ intersect the curve twice.
They satisfy the linear equivalence relations\\
\begin{align}
\begin{split}
[z_1]+[z_5]\sim&[z_2] \, ,\\
[z_3]-[z_0]\sim&[z_4] \, ,
\end{split}
\end{align}
and we see that there are two linearly independent two-sections.
There is no toric section and we cannot build a section out of toric divisors so we assume that \href{http://wwwth.mpp.mpg.de/members/jkeitel/weierstrass/data/122_0.txt}{(122,~0)} describes a genus-one fiber.
In particular we expect a $\mathbb{Z}_2$ discrete symmetry in the gauge theory.
As we explained above the geometry is self-dual so if the mirror conjecture is valid this fiber should also exhibit Mordell-Weil torsion.
Indeed the right hand side of the Weierstrass equation
\begin{align}
y^2=x^3+fx+g\,,
\end{align}
factors into a linear and a quadratic part and following \cite{Aspinwall:1998xj} this implies that the Jacobian has a $\mathbb{Z}_2$ torsion point.
In particular the Weierstrass coefficients calculated by\footnote{
We add here that one can use the line bundle associated to the divisor $D_3$, with
\begin{align}
z_3^2,\,z_0z_3z_4,\,z_0^2z_4^2,\,\frac{z_0z_1z_5^2z_8}{z_6}\in H^0\left(\mathbb{C},\mathcal{O}(2D_3)\right)\,,
\end{align}
to map the geometry into a quartic using
the procedure summarized in \cite{Braun:2014oya}. A map into a biquadric is possible via $u,v,w,z\in H^0\left(\mathbb{C},\mathcal{O}(D_2+D_5+D_8)\right)$, and the resulting equations take the simple form
\begin{align}
\begin{split}
q_1=&a_0u^2+a_1uw+a_3vw+a_2w^2+a_4wz\,,\\
q_2=&a_5v^2+a_6uz+a_7vz+a_9wz+a_8z^2\,,
\end{split}
\end{align}
in $\mathbb{P}^3$. This shows, that \href{http://wwwth.mpp.mpg.de/members/jkeitel/weierstrass/data/122_0.txt}{(122,~0)} is related to \href{http://wwwth.mpp.mpg.de/members/jkeitel/weierstrass/data/0_0.txt}{(0,~0)} by a conifold transition.
Blowing down all the exceptional divisors and deforming the complex structure into general position
merges the two linearly independent two-section into the single four-section of \href{http://wwwth.mpp.mpg.de/members/jkeitel/weierstrass/data/0_0.txt}{(0,~0)}.
A toric realization of this higgsing is given by the chain of blow-downs \href{http://wwwth.mpp.mpg.de/members/jkeitel/weierstrass/data/122_0.txt}{(122,~0)}$\rightarrow$\href{http://wwwth.mpp.mpg.de/members/jkeitel/weierstrass/data/62_1.txt}{(62,~1)}$\rightarrow$\href{http://wwwth.mpp.mpg.de/members/jkeitel/weierstrass/data/24_0.txt}{(24,~0)}$\rightarrow$\href{http://wwwth.mpp.mpg.de/members/jkeitel/weierstrass/data/6_1.txt}{(6,~1)}$\rightarrow$\href{http://wwwth.mpp.mpg.de/members/jkeitel/weierstrass/data/0_0.txt}{(0,~0)}.}
\cite{Braun:2014qka} can be written in the general form \cite{Aspinwall:1998xj}
\begin{align}
\begin{split}
f = A_4 - \frac13 A_2^2 \, ,&\quad g = \frac{1}{27} A_2 (2 A_2^2 - 9 A_4) \, ,\\
\Delta=& A_4^2 (4 A_4 - A_2^2)\,,
\end{split}
\end{align}
using the birational map
\begin{align}
\begin{split}
A_2 \, \rightarrow  \, & \phantom{=} 4 a_1 a_4 a_5 a_6 + a_3^2 a_6^2 - 2 a_1 a_3 a_6 a_7 + a_1^2 a_7^2\\
&- 4 a_0 a_2 a_7^2 - 4 a_1^2 a_5 a_8+16 a_0 a_2 a_5 a_8\\
&- 8 a_0 a_4 a_5 a_9 + 4 a_0 a_3 a_7 a_9 \, , \\
	A_4  \, \rightarrow \,  & \phantom{=}
 16 a_0 a_5 (a_4^2 a_5 - a_3 a_4 a_7 + a_3^2 a_8)\\
 &\cdot(a_2 a_6^2 - a_1 a_6 a_9 + a_0 a_9^2) \, .
\end{split}
\end{align}

In the following we want to compute the matter spectrum. The results are summarized in Table \ref{tab:spectrum122_0} and Table \ref{tab:1220Ideal}. 
From the Weierstrass data one deduces that the curve splits above the loci
\begin{align}
\begin{split}
L_0:\,a_0=0\, ,&\quad L_1:\,a_5=0 \, ,\\
L_2:\,a_4^2a_5-a_3&a_4a_7+a_3^2a_8=0 \, ,\\
L_3:\,a_2a_6^2-a_1&a_6a_9+a_0a_9^2=0 \, ,
\end{split}
\end{align}
into an $I_2$ fiber.

Above $L_0$ the section $p_1$ factors into $p^{(0)}_1\cdot z_0$ while above $L_1$ section $p_2$ factors into $p^{(1)}_2\cdot z_1$.
The two polynomial loci $L_2$ and $L_3$ each involve coefficients from both of the equations describing the complete intersection and are therefore of type $1$ from the classification in Appendix \ref{app:nonAbelian}.
We indeed find that the following linear combination factorizes above $L_2$
\begin{align}
a_3a_4z_0z_5p_2-(a_4a_5z_2+a_3a_8z_1z_5)p_1=p^{(2)}\cdot z_6
\label{eqn:l2combination}
\end{align}
with a polynomial $p^{(2)}$.
If $p_1=0$ then $z_6=0$ implies $p_2=0$ because from the Stanley-Reisner ideal it follows that neither $z_0$ nor $z_5$ can vanish simultaneously with $z_6$.
We therefore conclude that the singularity of the fiber above $L_2$ is torically resolved by the restriction of $[z_6]$ to the curve.
A similar combination factors above $L_3$ into
\begin{align}
a_6a_9z_1z_4p_1-(a_0a_9z_3+a_2a_6z_0z_4)p_2=p^{(3)}\cdot z_8 \, ,
\label{eqn:l3combination}
\end{align}
and the resolution divisor is given by the restriction of $[z_8]$ to the curve.
\begin{figure*}
\begin{minipage}{.23\textwidth}
\includegraphics[width=\linewidth]{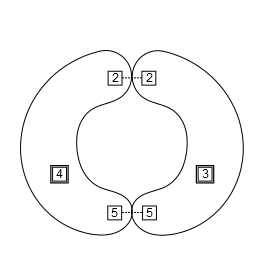}
\begin{tikzpicture}[remember picture,overlay,node distance=4mm, >=latex',block/.style = {draw, rectangle, minimum height=65mm, minimum width=83mm,align=center},]
\node[overlay,align=center] at (2,0) {Gauge locus $L_0$,\\{\small$a_0=0$}};
\end{tikzpicture}
\end{minipage}
\begin{minipage}{.23\textwidth}
\includegraphics[width=\linewidth]{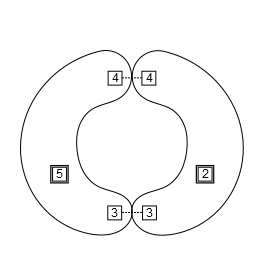}
\begin{tikzpicture}[remember picture,overlay,node distance=4mm, >=latex',block/.style = {draw, rectangle, minimum height=65mm, minimum width=83mm,align=center},]
\node[overlay,align=center] at (2,0) {Gauge locus $L_1$,\\{\small$a_5=0$}};
\end{tikzpicture}
\end{minipage}
\begin{minipage}{.23\textwidth}
\includegraphics[width=\linewidth]{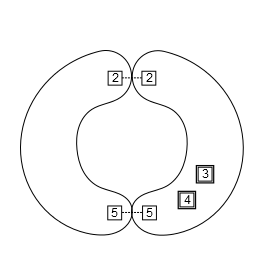}
\begin{tikzpicture}[remember picture,overlay,node distance=4mm, >=latex',block/.style = {draw, rectangle, minimum height=65mm, minimum width=83mm,align=center},]
\node[overlay,align=center] at (2,0) {Gauge locus $L_2$,\\{\small$a_4^2a_5-a_3a_4a_7+a_3^2a_8=0$}};
\end{tikzpicture}
\end{minipage}
\begin{minipage}{.23\textwidth}
\includegraphics[width=\linewidth]{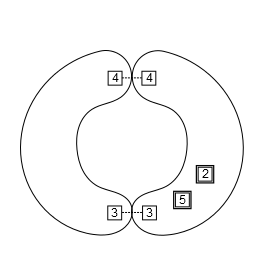}
\begin{tikzpicture}[remember picture,overlay,node distance=4mm, >=latex',block/.style = {draw, rectangle, minimum height=65mm, minimum width=83mm,align=center},]
\node[overlay,align=center] at (2,0) {Gauge locus $L_3$,\\{\small$a_2a_6^2-a_1a_6a_9+a_0a_9^2=0$}};
\end{tikzpicture}
\end{minipage}
\vspace{.5cm}
\caption{Splitting of the fiber above loci of codimension one. A number $i$ in a box means that the two-section $[z_i]$ intersects that component.
A double-box is used when the two-section intersects that component twice.}
\label{fig:codim1Curve}
\end{figure*}
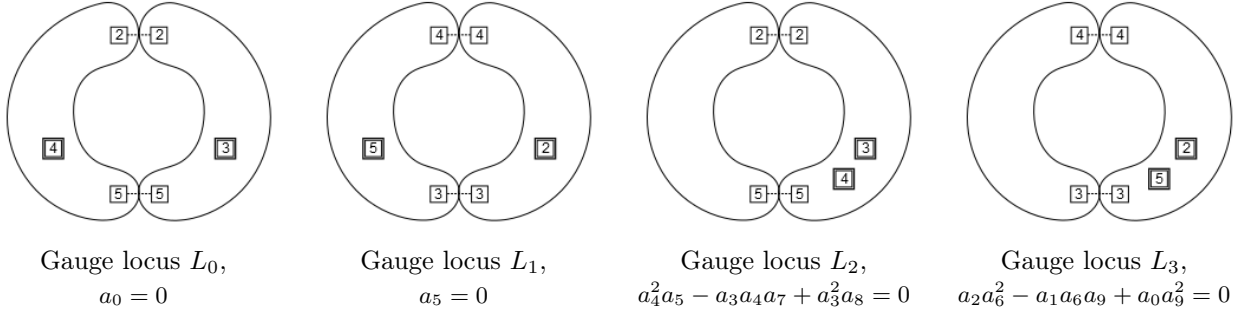

There are four vertices that correspond to divisors which intersect the curve
\begin{align}
\begin{split}
z_2&=(0,\,0,\,1) \, ,\quad z_3=(-1,\,-1,\,-1)\, ,\\
z_4&=(1,\,-1,\,-1) \, ,\quad z_5=(0,\,2,\,1) \, .
\end{split}
\end{align}
One can directly see that the vertices span a sublattice of finite index in $\mathbb{Z}^3$ and that the dual lattice is generated by
\begin{align}
b_1=\left(\frac{1}{2},\frac{1}{2},0\right)\, ,\quad b_2=(0,1,0)\, , \quad b_3=(0,0,1)\, .
\end{align}
The element $b_1$ corresponds to a trivial $\mathbb{Q}$-divisor
\begin{align}
[z_5]-[z_3]+\frac{1}{2}[z_0]+\frac{1}{2}[z_1]-\frac{1}{2}[z_6]+\frac{1}{2}[z_8] \, .
\label{eqn:122_0Shioda}
\end{align}
As was noted in \cite{Mayrhofer:2014opa} this implies that
\begin{align}
[z_5]-[z_3]\sim-\left(\frac{1}{2}[z_0]+\frac{1}{2}[z_1]-\frac{1}{2}[z_6]+\frac{1}{2}[z_8]\right) \, ,
\end{align}
and therefore the right hand side of this equation has integral intersection with all curves.
Although $[z_5]$ and $[z_3]$ both intersect the curve twice we interpret (\ref{eqn:122_0Shioda}) as
the ambient part of a generalized torsion Shioda map associated to the multi-sections.

The intersection patterns of the multi-sections with the singular fibers at codimension one are shown in Figure~\ref{fig:codim1Curve}
and it is a priori not clear which components should be taken as the affine nodes.
However we guess that in the complete fibration the torsion Shioda map takes the form
\begin{align}
\begin{split}
\sigma^{(2)}_T=&[z_5]-[z_3]-\pi^{-1}\left(\bar{K}_B\right)\\
&+\frac{1}{2}[z_0]+\frac{1}{2}[z_1]+\frac{1}{2}\left([L_2]-[z_6]\right)+\frac{1}{2}[z_8] \, .
\end{split}
\label{eqn:122_0fullTorsionShioda}
\end{align}
This constrains the weights $(w_1,\,w_2,\,w_3,\,w_4)$ of the representations that appear in the matter-spectrum to satisfy
\begin{align}
\label{eq:evenweights}
w_1+w_2+w_3+w_4\in2\mathbb{Z} \, ,
\end{align}
which in particular excludes fundamental matter. 

From the two-sections we can also write down a discrete Shioda map which determines the charge assignment of the matter curves under a discrete symmetry.
As the two-sections intersect the codimension one fibers, we have
to orthogonalize the discrete Shioda map appropriately with respect to the $SU(2)^4$ resolution divisors $[z_i]$ with $i=0,1,6,8$.
We obtain the divisors
\begin{subequations}
\begin{align}
\sigma^{(2)}_{d,2} =& [z_2] + \frac12 \left( [z_0] + [z_8]  \right) \, , \\ 
\sigma^{(2)}_{d,3} =& [z_3] + \frac12 \left( [z_1]+ [z_6]  \right) \, , \\ 
\sigma^{(2)}_{d,4} =& [z_4] + \frac12 \left( [z_1] + [z_6]  \right) \, , \\ 
\sigma^{(2)}_{d,5} =& [z_5] + \frac12 \left( [z_0] + [z_8] \right) \, ,
\end{align}
\end{subequations}
which are all equivalent up to integral multiples of resolution divisors.
The orthogonalization procedure leads to half-integer charges and produces an effective enhancement of the $\mathbb{Z}_2$ symmetry to $\mathbb{Z}_4$ by
the SU(2) center elements as was already observed in the $\mathbb{P}_{112}$ case \cite{Klevers:2014bqa,Garcia-Etxebarria:2014qua}.
In the following we will use $\sigma^{(2)}_{d,4}$ as the charge operator.
\vspace{1cm}
\subsection*{The matter spectrum}
The analysis of the matter spectrum is slightly more involved than in the hypersurface case
and we describe it for three of the loci in detail.
\begin{table}
{ \small
\begin{tabular}{c|c|c}
Locus & $(f,g,\Delta)$ & $SU(2)^4$ Rep. \\ \hline \hline
$a_0=0$ & $(0,0,2)$ & (\three,\one,\one,\one) \\ \hline
$a_5=0$ & $(0,0,2)$ & (\one,\three,\one,\one) \\ \hline
$a_4^2 a_5 - a_3 a_4 a_7  + a_3^2 a_8 = 0$ & $(0,0,2)$ & (\one,\one,\three,\one) \\ \hline
$a_2 a_6^2 - a_1 a_6 a_9 + a_0 a_9^2 = 0$ & $(0,0,2)$ & (\one,\one,\one,\three)
\end{tabular}
}
\caption{Codimension one loci of nef \href{http://wwwth.mpp.mpg.de/members/jkeitel/weierstrass/data/122_0.txt}{(122,~0)}.}
\label{tab:spectrum122_0}
\end{table}
There are eight codimension two loci at which the splitting of the fiber is enhanced.
At each of those the vanishing order of $(f,\,g,\,\Delta)$ is enhanced to $(0,\,0,\,4)$.
\begin{itemize}
\item \textbf{Matter locus 1:} $a_0=0,\,a_5=0$\, .\\
At this locus the equations split as
\begin{align}
\begin{split}
p_1^{(1)}= p'^{(1)}_1\cdot z_0 \, ,\\
p_2^{(1)}= p'^{(1)}_2\cdot z_1\, .
\end{split}
\end{align}
and in the patch with variables $z_0,\,z_1$ and $z_4$ we find
\begin{align}
\begin{split}
p'^{(1)}_1=a_3+a_4z_1+a_1z_4+a_2z_0z_4^2\, ,\\
p'^{(1)}_2=a_7+a_8z_1+a_6z_4+a_9z_0z_4^2 \, ,
\end{split}
\end{align}
The fiber splits into the four components
\begin{align}
\begin{split}
\mathbb{P}_1^{1,1}=\{p'^{(1)}_1=0\}\cap \{z_0=0\}\, ,\\\quad\mathbb{P}_1^{1,2}=\{p'^{(1)}_2=0\}\cap \{z_1=0\}\, ,\\
\mathbb{P}_1^{1,3}=\{z_0=0\}\cap \{z_1=0\}\, ,\\\quad\mathbb{P}_1^{1,4}=\{p'^{(1)}_1=0\}\cap \{p'^{(1)}_2=0\}\, ,
\end{split}
\end{align}
and we calculate the intersections shown in Figure~\ref{fig:codim2Curve}.
This leads to matter in the bifundamental representation $(\two,\two,\one,\one)$ with highest weight $(1,1,0,0)$.
\item \textbf{Matter locus 2:} $a_0=0\, ,\,a_6=0$\, .\\
At this locus we can go to the patch with variables $z_0,\,z_2,\,z_8$ setting all other variables to one and find
\begin{align}
\begin{split}
p_1^{(2)}=&p_1'^{(2)}\cdot z_0\, ,\\
p_2^{(2)}=&p_2'^{(2)}\cdot z_8+a_9z_0 \, .
\end{split}
\end{align}
with
\begin{align}
\begin{split}
p_1'^{(2)}=&a_1+a_2z_0+a_4z_8+a_3z_2z_8\, ,\\
p_2'^{(2)}=&a_5z_2^2+a_7z_2+a_8\, .
\end{split}
\end{align}
We see that the fiber is non-split and find only three components
{\small
\begin{align}
\mathbb{P}_1^{2,1}=&\{a_5z_2^2+a_7z_1z_2z_5+a_8z_1^2z_5^2=0\}\cap\{z_0=0\}\, , \nonumber \\
\mathbb{P}_{1}^{2,2}=&\{z_8=0\}\cap\{z_0=0\}\, ,\\
\mathbb{P}_1^{2,3}=&\overline{(\{p_1=0\}\cap\{p_2=0\})\backslash(\mathbb{P}_{1}^{2,1}\cup\mathbb{P}_{1}^{2,2})}\, \nonumber ,
\end{align}
}
with intersection numbers
\begin{align}
\begin{split}
\#\left(\mathbb{P}_1^{2,1}\cap\mathbb{P}_1^{2,2}\right)=2\, ,\quad &\#\left(\mathbb{P}_1^{2,1}\cap\mathbb{P}_1^{2,3}\right)=2\, ,\\
\#\left(\mathbb{P}_1^{2,2}\right . &\cap \left . \mathbb{P}_1^{2,3}\right)=0\, .
\end{split}
\end{align}
The component $\mathbb{P}_1^{2,1}=\mathbb{P}_{1,+}^{2,1}\cup \mathbb{P}_{1,-}^{2,1}$ can be factored over a field extension into
{
\small
\begin{align}
\left\{z_2=\frac{1}{2a_5}\left(a_7\pm\sqrt{a_7^2-4a_5a_8}\right)\right\}\cap \{z_0=0\}\, ,
\end{align}
}
and the two components are exchanged under a monodromy around the codimension three locus
\begin{align}
a_0=a_6=a_7^2-4a_5a_8=0\,.
\end{align}
The components experiencing monodromy are drawn with dashed lines  in Figure~\ref{fig:codim2Curve}
and again we find bifundamental matter, now in the representation $(\two,\one,\one,\two)$ with highest weight $(1,0,0,1)$.
\item\textbf{Matter locus 3:} $a_0=0 \, ,\,a_2a_6-a_1a_9=0\, .$\\
From the previous discussion we know that $z_0=p_2=0$ will again be a component above this locus.
We also expect $[z_8]$ to give a component and indeed imposing $a_0=z_8=0$ the equations become
\begin{align}
\begin{split}
(a_1z_3+a_2z_0z_4)\cdot z_0z_4z_6=0 \, ,\\
(a_6z_3+a_9z_0z_4)\cdot z_0z_4z_6=0\, .
\end{split}
\end{align}
It follows from the Stanley-Reisner ideal that $z_8$ cannot vanish simultaneously with $z_0,\,z_4$ or $z_6$.
Using $a_2a_6-a_1a_9=0$ we find the second component $(a_1z_3+a_2z_0z_4)=z_8=0$ above this matter locus.
Again the last component is non-split and we have three components
{\small
\begin{align}
\begin{split}
\mathbb{P}_1^{3,1}=&\{p_2=0\}\cap\{z_0=0\} \, ,\\
\mathbb{P}_1^{3,2}=&\{a_6z_3+a_9z_0z_4=0\}\cap\{z_8=0\} \, ,\\
\mathbb{P}_1^{3,3}=&\overline{(\{p_1=0\}\cap\{p_2=0\})\backslash(\mathbb{P}_{1}^{3,1}\cup\mathbb{P}_{1}^{3,2})}\, .
\end{split}
\end{align}
}
This time $\mathbb{P}_1^{3,3}=\mathbb{P}_{1,+}^{3,3}\cup \mathbb{P}_{1,-}^{3,3}$ factors over a field extension and the two components are exchanged under monodromy.
The splitting and intersections are shown in Figure~\ref{fig:codim2Curve} and we find the same representation as above, i.e. $(\two,\one,\one,\two)$ with highest weight $(1,0,0,1)$.
\end{itemize}
The splitting at the remaining matter loci can be determined by similar calculations.
We find the representations listed in Table \ref{tab:1220Splittings} and the fiber splittings are shown in Figure~\ref{fig:codim2Curve}.
In the picture we also mark the intersections of the multi-sections with the matter curves.
\begin{table*}
{ \small\renewcommand{\arraystretch}{1.5}
\begin{tabular}{|c|c|c|c|} \hline
Locus & $(f,g,\Delta)$ & $SU(2)^4\times \mathbb{Z}_4 $ Rep. & Fiber Components \\ \hline \hline
$\begin{array}{r}
a_0 = 0\, ,\,a_5 = 0 
\end{array}$
& $(0,0,4)$ & (\two,\two,\one,\one)$_{\frac12}$\phantom{$^\prime$}&
$\begin{array}{c}
\mathbb{P}_1^{1,1}=\overline{(\{p_2=0\}\backslash\{z_1=0\})}\cap\{z_0=0\}\\
\mathbb{P}_1^{1,2}=\overline{(\{p_1=0\}\backslash\{z_0=0\})}\cap\{z_1=0\}\\
\mathbb{P}_1^{1,3}=\{z_0=0\}\cap\{z_1=0\}\\
\mathbb{P}_1^{1,4}=\overline{(\{p_1=0\}\cap\{p_2=0\})\backslash(\mathbb{P}_1^{1,1}\cup\mathbb{P}_1^{1,2}\cup\mathbb{P}_1^{1,3})}
\end{array}$\\ \hline
$\begin{array}{r}
a_0 = 0\, ,\,a_6 = 0 
\end{array}$
& $(0,0,4)$ & (\two,\one,\one,\two)$_{1}$\phantom{$^\prime$}&
$\begin{array}{c}
\mathbb{P}_{1,+}^{2,1}\cup \mathbb{P}_{1,-}^{2,1}=\{a_5z_2^2+a_7z_1z_2z_5+a_8z_1^2z_5^2\}\cap\{z_0=0\}\\
\mathbb{P}_{1}^{2,2}=\{z_8=0\}\cap\{z_0=0\}\\
\mathbb{P}_1^{2,3}=\overline{(\{p_1=0\}\cap\{p_2=0\})\backslash(\mathbb{P}_{1,+}^{2,1}\cup \mathbb{P}_{1,-}^{2,1}\cup\mathbb{P}_{1}^{2,2})}
\end{array}$
\\ \hline
$\begin{array}{c}
a_0 = 0 \, , \\  a_2 a_6 - a_1 a_9 = 0 
\end{array}$
& $(0,0,4)$ & (\two,\one,\one,\two)$_{0}^\prime$ &
$
\begin{array}{c}
\mathbb{P}_1^{3,1}=\{p_2=0\}\cap\{z_0=0\}\\
\mathbb{P}_1^{3,2}=\{a_6z_3+a_9z_0z_4=0\}\cap\{z_8=0\}\\
\mathbb{P}_{1,+}^{3,3}\cup\mathbb{P}_{1,-}^{3,3}=\overline{(\{p_1=0\}\cap\{p_2=0\})\backslash(\mathbb{P}_{1}^{3,1}\cup\mathbb{P}_{1}^{3,2})}
\end{array}
$
\\ \hline
$\begin{array}{r}
a_5 = 0\, ,\,a_3 = 0 
\end{array}$
& $(0,0,4)$ & (\one,\two,\two,\one)$_0$\phantom{$^\prime$}&
$
\begin{array}{c}
\mathbb{P}_1^{4,1}=\{z_1=0\}\cap\{z_6=0\}\\
\mathbb{P}_{1,+}^{4,2}\cup\mathbb{P}_{1,-}^{4,2}=\{a_0z_3^2+a_0z_0z_3z_4+a_2z_0^2z_4^2=0\}\cap\{z_1=0\}\\
\mathbb{P}_1^{4,3}=\overline{(\{p_1=0\}\cap\{p_2=0\})\backslash(\mathbb{P}_1^{4,1}\cup\mathbb{P}_{1,+}^{4,2}\cup\mathbb{P}_{1,-}^{4,2})}
\end{array}
$
\\ \hline
$\begin{array}{c}
a_5 = 0\, , \\ a_4 a_7 - a_3 a_8 = 0 
\end{array}$
& $(0,0,4)$ & (\one,\two,\two,\one)$_1^\prime$ &
$
\begin{array}{c}
\mathbb{P}_1^{5,1}=\{p_1=0\}\cap\{z_1=0\}\\
\mathbb{P}_1^{5,2}=\{a_7z_2+a_8z_1z_5=0\}\cap\{z_6=0\}\\
\mathbb{P}_{1,+}^{5,3}\cup\mathbb{P}_{1,-}^{5,3}=\overline{(\{p_1=0\}\cup\{\mathbb{P}_1^{8,1}p_2=0\})\backslash(\mathbb{P}_1^{5,1}\cup\mathbb{P}_1^{5,2})}
\end{array}
$
\\ \hline
$\begin{array}{c}
a_0 = 0 \, , \\ a_4^2 a_5 - a_3 a_4 a_7  + a_3^2 a_8 = 0
\end{array}$
& $(0,0,4)$ & (\two,\one,\two,\one)$_{-\frac12}$\phantom{$^\prime$}&
$
\begin{array}{c}
\mathbb{P}_1^{6,1}=\{a_3z_2+a_4z_1z_5=0\}\cap\{z_4=0\}\\
\mathbb{P}_1^{6,2}=\{a_3z_2+a_4z_1z_5=0\}\cap\{z_6=0\}\\
\mathbb{P}_1^{6,3}=\{p_2=0\}\cap\{z_0=0\}\\
\mathbb{P}_1^{6,4}=\overline{(\{p_1=0\}\cup\{p_2=0\})\backslash(\mathbb{P}_1^{6,1}\cup\mathbb{P}_1^{6,2}\cup\mathbb{P}_1^{6,3})}
\end{array}
$
\\ \hline
$\begin{array}{c}
a_5 = 0 \, , \\a_2 a_6^2 - a_1 a_6 a_9 + 
   a_0 a_9^2 = 0 
\end{array}$
& $(0,0,4)$ & (\one,\two,\one,\two)$_{\frac12}$\phantom{$^\prime$}&
$
\begin{array}{c}
\mathbb{P}_1^{7,1}=\{a_6z_3+a_9z_0z_4=0\}\cap\{z_5=0\}\\
\mathbb{P}_1^{7,2}=\{a_6z_3+a_9z_0z_4=0\}\cap\{z_8=0\}\\
\mathbb{P}_1^{7,3}=\{p_1=0\}\cap\{z_1=0\}\\
\mathbb{P}_1^{7,4}=\overline{(\{p_1=0\}\cup\{p_2=0\})\backslash(\mathbb{P}_1^{7,1}\cup\mathbb{P}_1^{7,2}\cup\mathbb{P}_1^{7,3})}
\end{array}
$
\\ \hline
$\begin{array}{c}
a_4^2 a_5 - a_3 a_4 a_7 + a_3^2 a_8 =0\, , \\ a_2 a_6^2 - a_1 a_6 a_9 + 
   a_0 a_9^2 = 0
\end{array}$
& $(0,0,4)$ & (\one,\one,\two,\two)$_{\frac12}$\phantom{$^\prime$}&
$
\begin{array}{c}
\mathbb{P}_1^{8,1}=\{a_3z_2+a_4z_1z_5=0\}\cap\{a_6z_3+a_9z_0z_4=0\}\\
\mathbb{P}_1^{8,2}=\{a_3z_2+a_4z_1z_5=0\}\cap\{z_6=0\}\\
\mathbb{P}_1^{8,3}=\{a_6z_3+a_9z_0z_4=0\}\cap\{z_8=0\}\\
\mathbb{P}_1^{8,4}=\overline{(\{p_1=0\}\cap\{p_2=0\})\backslash(\mathbb{P}_1^{8,1}\cup\mathbb{P}_1^{8,2}\cup\mathbb{P}_1^{8,3})}
\end{array}
$
\\  \hline
\end{tabular}
}
\caption{\label{tab:1220Splittings}The vanishing order of the discriminant, components of the fiber and matter representations above codimension two loci with enhanced singularities.
A line over a component denotes the closure in the Zariski topology, e.g. $\overline{\{(az_3+b z_0z_4)\cdot z_1=0\}\backslash\{z_1=0\}}=\{az_3+b z_0z_4=0\}$.}
\label{tab:1220Ideal}
\end{table*}
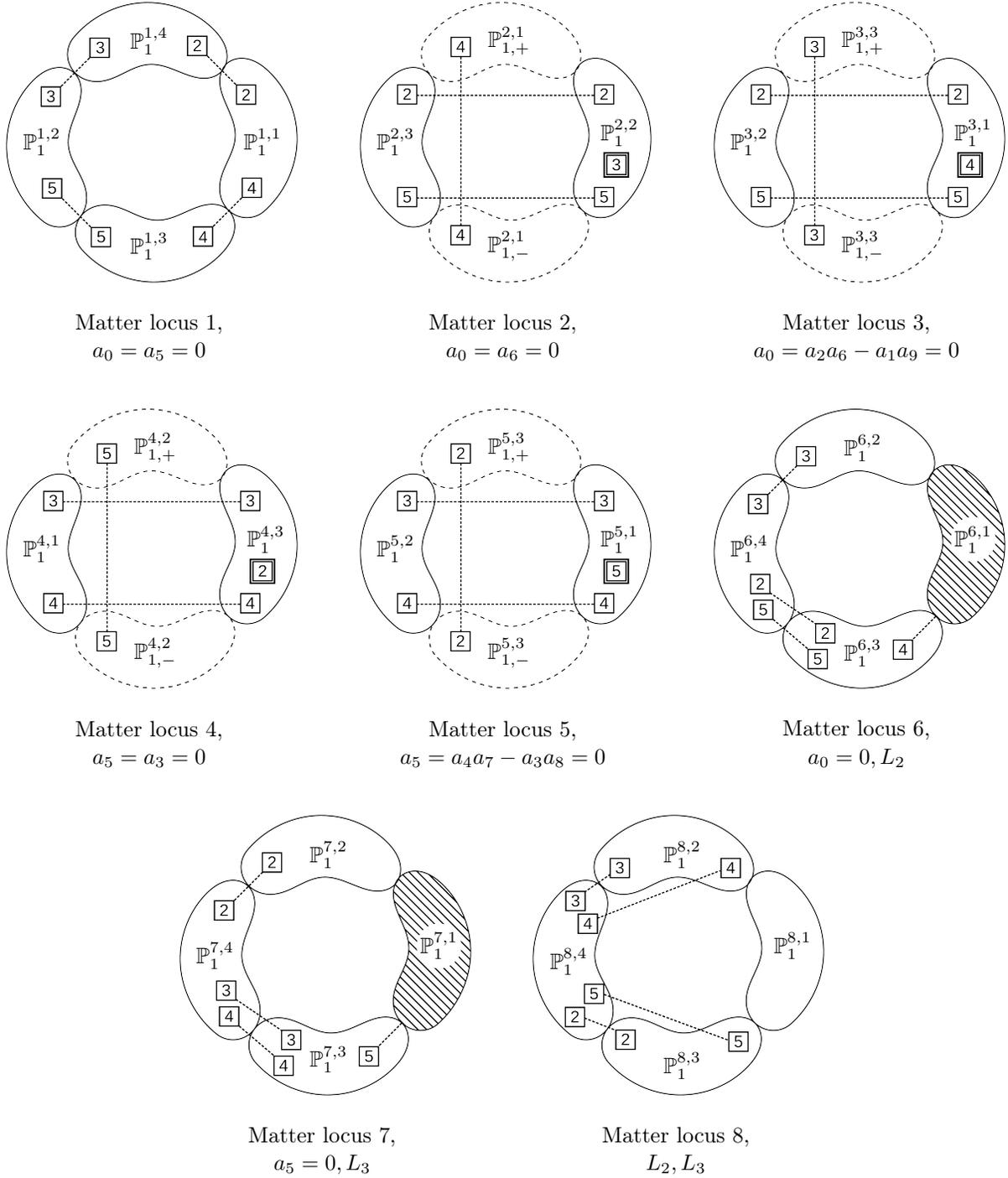
\begin{figure*}
\centering
\begin{minipage}{.305\textwidth}
\includegraphics[width=\linewidth]{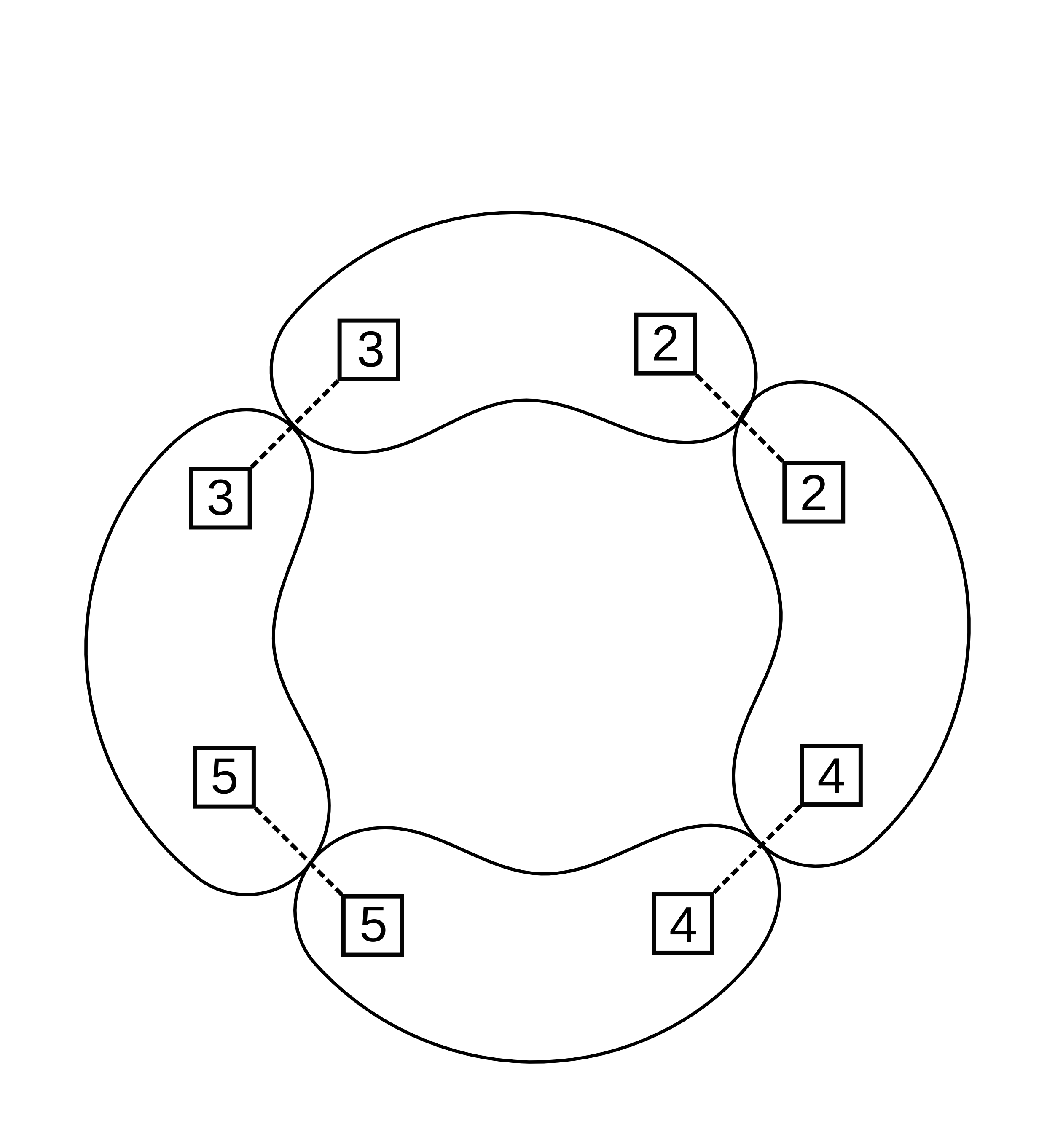}
\begin{tikzpicture}[remember picture,overlay,node distance=4mm, >=latex',block/.style = {draw, rectangle, minimum height=65mm, minimum width=83mm,align=center},]
\node[overlay,align=center] at (2.7,4.7) {$\mathbb{P}_1^{1,4}$};
\node[overlay,align=center] at (4.5,3.1) {$\mathbb{P}_1^{1,1}$};
\node[overlay,align=center] at (1,3.1) {$\mathbb{P}_1^{1,2}$};
\node[overlay,align=center] at (2.7,1.45) {$\mathbb{P}_1^{1,3}$};
\node[overlay,align=center] at (2.7,0) {Matter locus 1,\\$a_0=a_5=0$};
\end{tikzpicture}
\end{minipage}
\begin{minipage}{.305\textwidth}
\includegraphics[width=\linewidth]{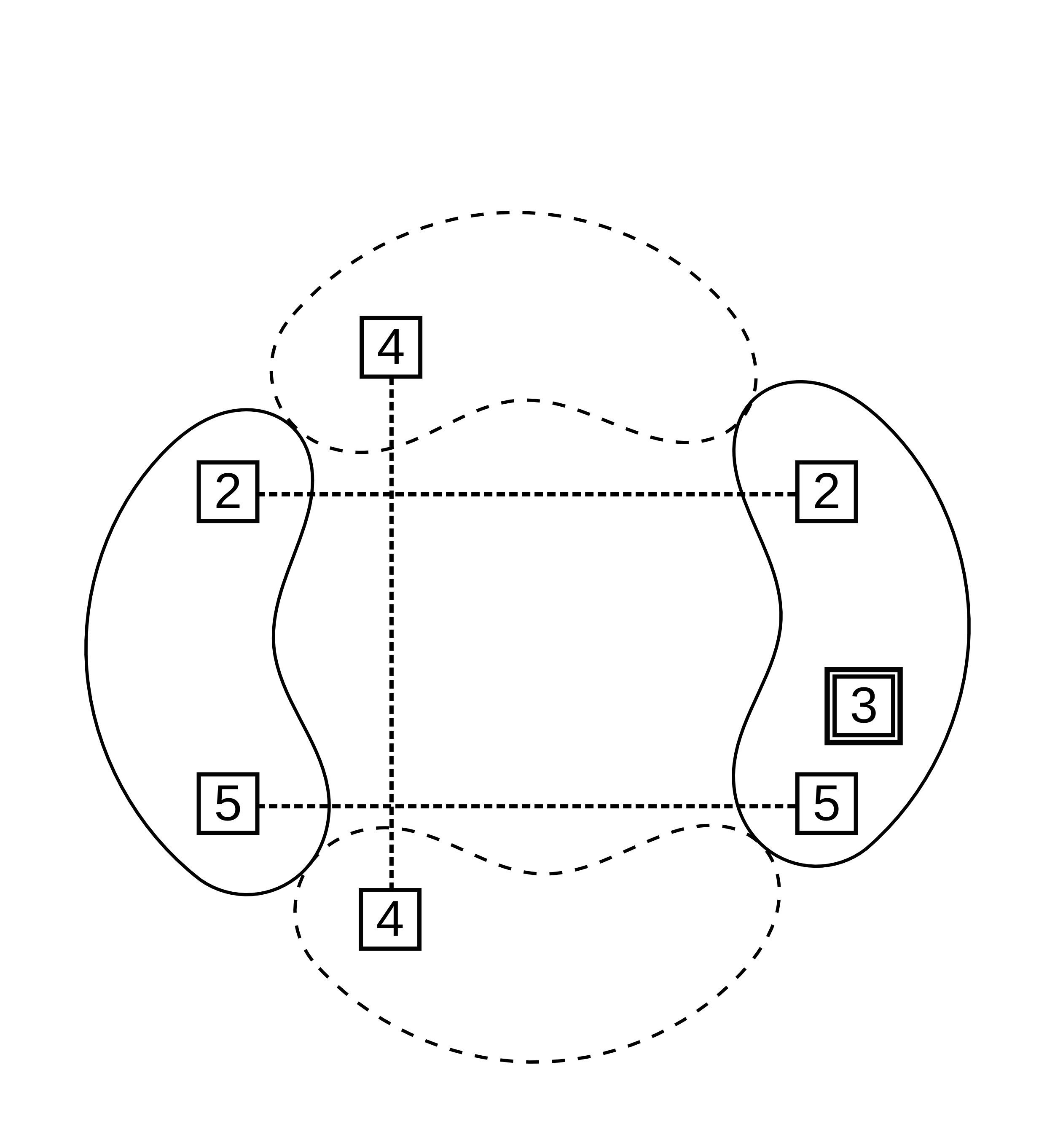}
\begin{tikzpicture}[remember picture,overlay,node distance=4mm, >=latex',block/.style = {draw, rectangle, minimum height=65mm, minimum width=83mm,align=center},]
\node[overlay,align=center] at (2.7,0) {Matter locus 2,\\$a_0=a_6=0$};
\node[overlay,align=center] at (2.8,4.7) {$\mathbb{P}_{1,+}^{2,1}$};
\node[overlay,align=center] at (4.55,3.25) {$\mathbb{P}_1^{2,2}$};
\node[overlay,align=center] at (1,3.1) {$\mathbb{P}_1^{2,3}$};
\node[overlay,align=center] at (2.8,1.45) {$\mathbb{P}_{1,-}^{2,1}$};
\end{tikzpicture}
\end{minipage}
\begin{minipage}{.305\textwidth}
\includegraphics[width=\linewidth]{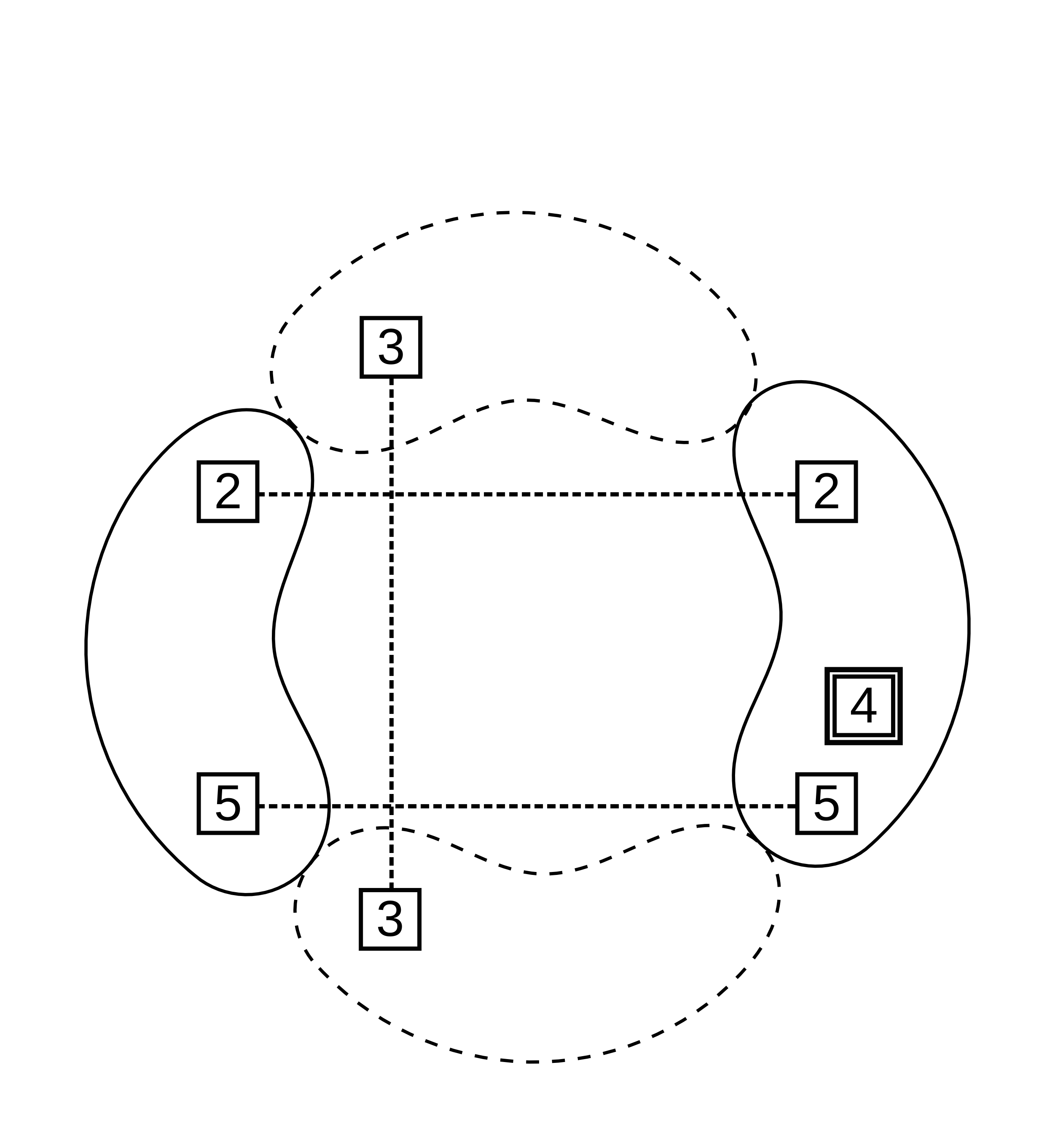}
\begin{tikzpicture}[remember picture,overlay,node distance=4mm, >=latex',block/.style = {draw, rectangle, minimum height=65mm, minimum width=83mm,align=center},]
\node[overlay,align=center] at (2.7,0) {Matter locus 3,\\$a_0=a_2a_6-a_1a_9=0$};
\node[overlay,align=center] at (2.8,4.7) {$\mathbb{P}_{1,+}^{3,3}$};
\node[overlay,align=center] at (4.55,3.25) {$\mathbb{P}_1^{3,1}$};
\node[overlay,align=center] at (1,3.1) {$\mathbb{P}_1^{3,2}$};
\node[overlay,align=center] at (2.8,1.45) {$\mathbb{P}_{1,-}^{3,3}$};
\end{tikzpicture}
\end{minipage}
\begin{minipage}{.305\textwidth}
\includegraphics[width=\linewidth]{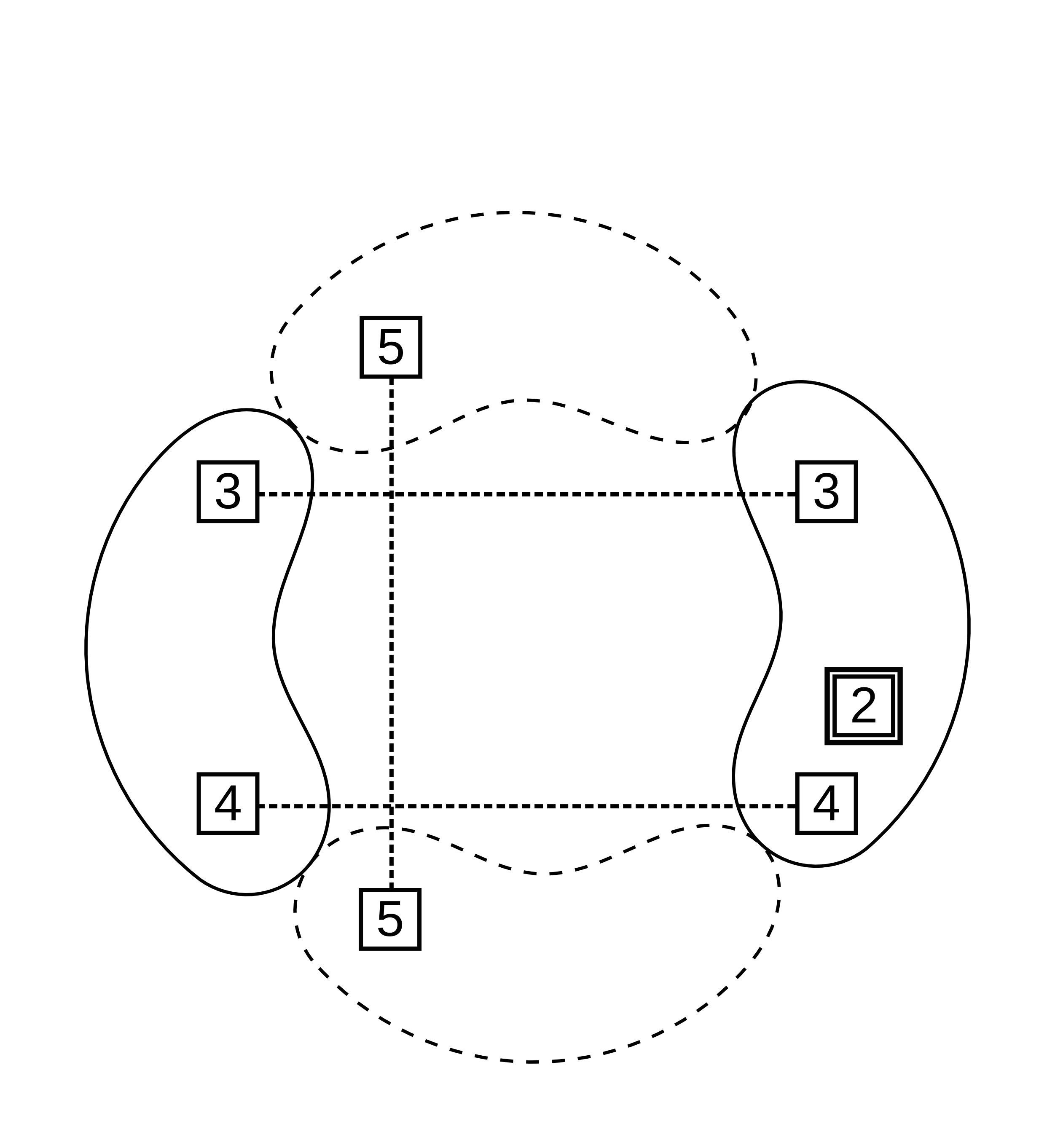}
\begin{tikzpicture}[remember picture,overlay,node distance=4mm, >=latex',block/.style = {draw, rectangle, minimum height=65mm, minimum width=83mm,align=center},]
\node[overlay,align=center] at (2.7,0) {Matter locus 4,\\$a_5=a_3=0$};
\node[overlay,align=center] at (2.8,4.7) {$\mathbb{P}_{1,+}^{4,2}$};
\node[overlay,align=center] at (4.55,3.25) {$\mathbb{P}_1^{4,3}$};
\node[overlay,align=center] at (1,3.1) {$\mathbb{P}_1^{4,1}$};
\node[overlay,align=center] at (2.8,1.45) {$\mathbb{P}_{1,-}^{4,2}$};
\end{tikzpicture}
\end{minipage}
\begin{minipage}{.305\textwidth}
\includegraphics[width=\linewidth]{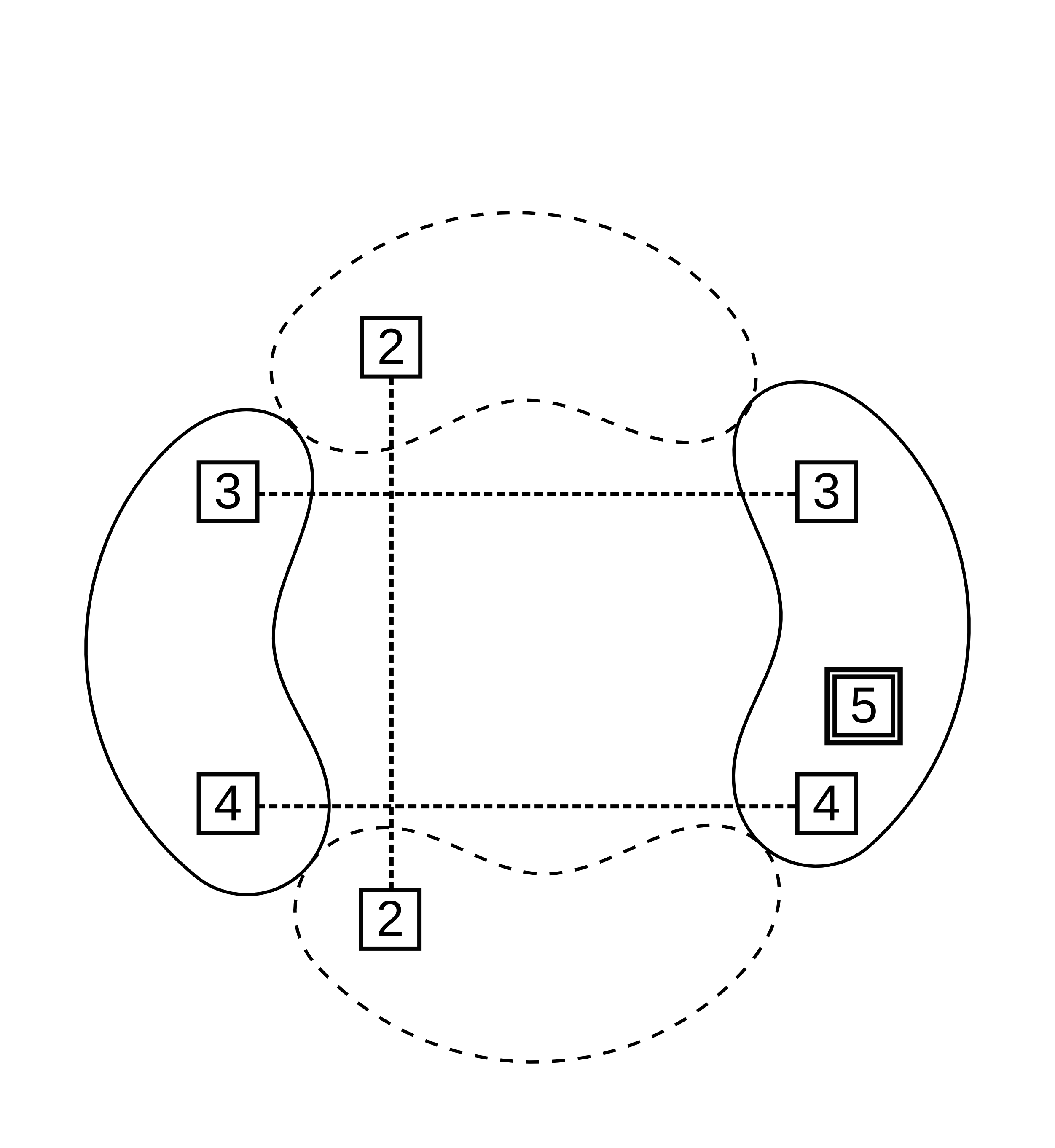}
\begin{tikzpicture}[remember picture,overlay,node distance=4mm, >=latex',block/.style = {draw, rectangle, minimum height=65mm, minimum width=83mm,align=center},]
\node[overlay,align=center] at (2.7,0) {Matter locus 5,\\$a_5=a_4a_7-a_3a_8=0$};
\node[overlay,align=center] at (2.8,4.7) {$\mathbb{P}_{1,+}^{5,3}$};
\node[overlay,align=center] at (4.55,3.25) {$\mathbb{P}_1^{5,1}$};
\node[overlay,align=center] at (1,3.1) {$\mathbb{P}_1^{5,2}$};
\node[overlay,align=center] at (2.8,1.45) {$\mathbb{P}_{1,-}^{5,3}$};
\end{tikzpicture}
\end{minipage}
\begin{minipage}{.305\textwidth}
\includegraphics[width=\linewidth]{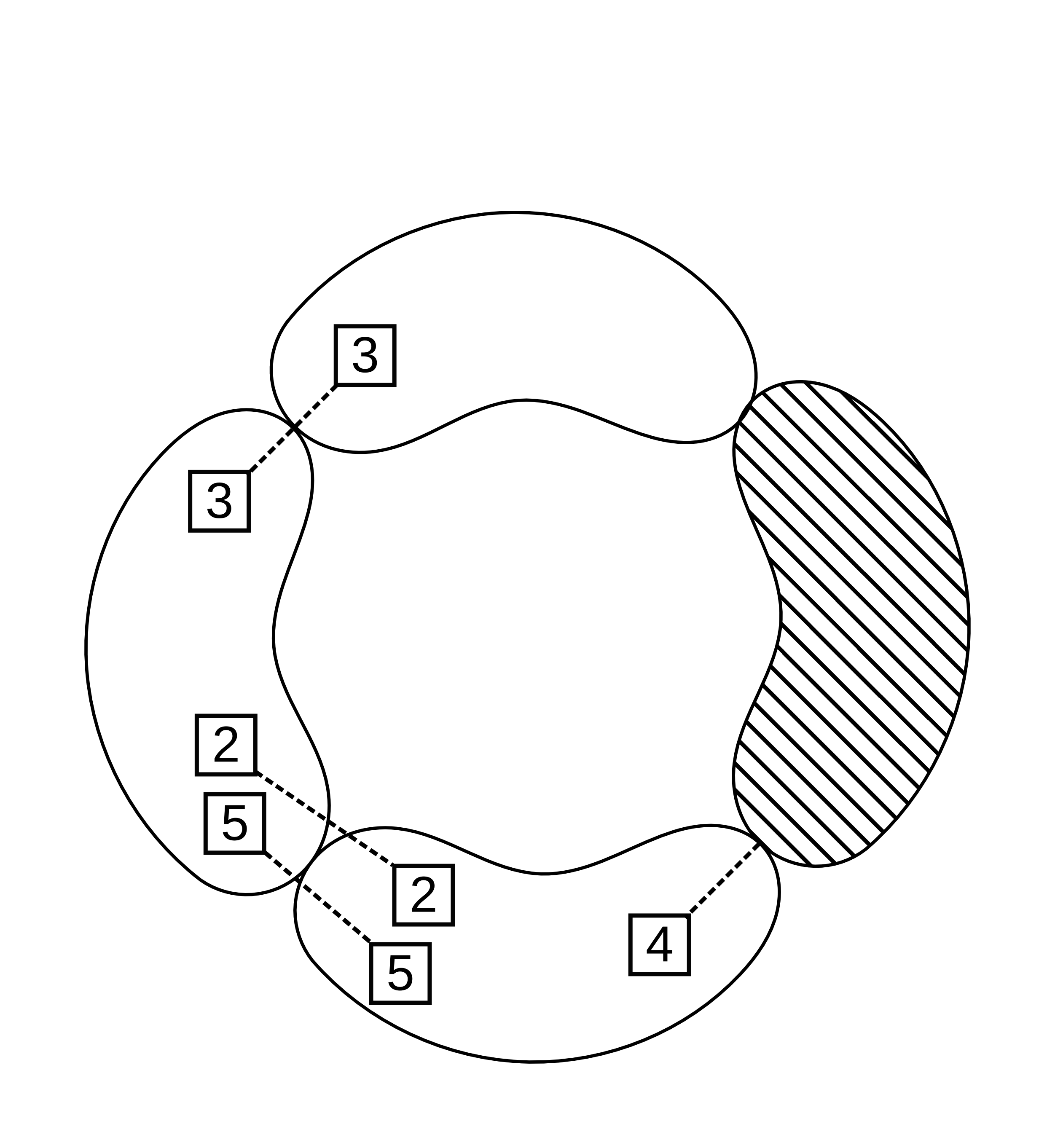}
\begin{tikzpicture}[remember picture,overlay,node distance=4mm, >=latex',block/.style = {draw, rectangle, minimum height=65mm, minimum width=83mm,align=center},]
\node[overlay,align=center] at (2.7,0) {Matter locus 6,\\$a_0=0, L_2$};
\draw[white,fill=white] (4.55,3.25) circle (.35cm);
\node[overlay,align=center] at (2.8,4.7) {$\mathbb{P}_{1}^{6,2}$};
\node[overlay,align=center] at (4.55,3.25) {$\mathbb{P}_1^{6,1}$};
\node[overlay,align=center] at (1,3.1) {$\mathbb{P}_1^{6,4}$};
\node[overlay,align=center] at (2.8,1.45) {$\mathbb{P}_{1}^{6,3}$};
\end{tikzpicture}
\end{minipage}
\begin{minipage}{.305\textwidth}
\includegraphics[width=\linewidth]{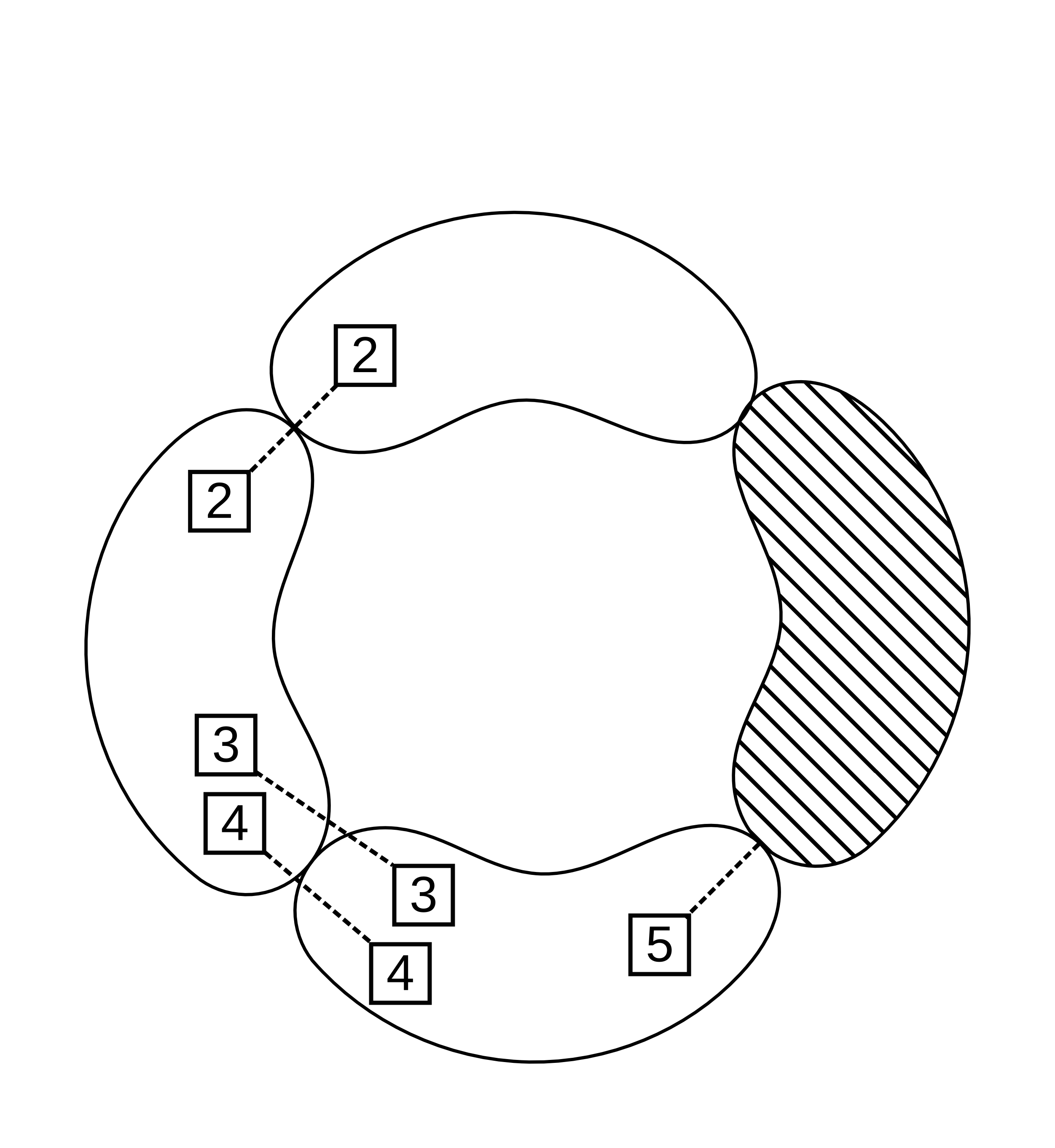}
\begin{tikzpicture}[remember picture,overlay,node distance=4mm, >=latex',block/.style = {draw, rectangle, minimum height=65mm, minimum width=83mm,align=center},]
\node[overlay,align=center] at (2.7,0) {Matter locus 7,\\$a_5=0, L_3$};
\draw[white,fill=white] (4.55,3.25) circle (.35cm);
\node[overlay,align=center] at (2.8,4.7) {$\mathbb{P}_{1}^{7,2}$};
\node[overlay,align=center] at (4.55,3.25) {$\mathbb{P}_1^{7,1}$};
\node[overlay,align=center] at (1,3.1) {$\mathbb{P}_1^{7,4}$};
\node[overlay,align=center] at (2.8,1.45) {$\mathbb{P}_{1}^{7,3}$};
\end{tikzpicture}
\end{minipage}
\begin{minipage}{.305\textwidth}
\includegraphics[width=\linewidth]{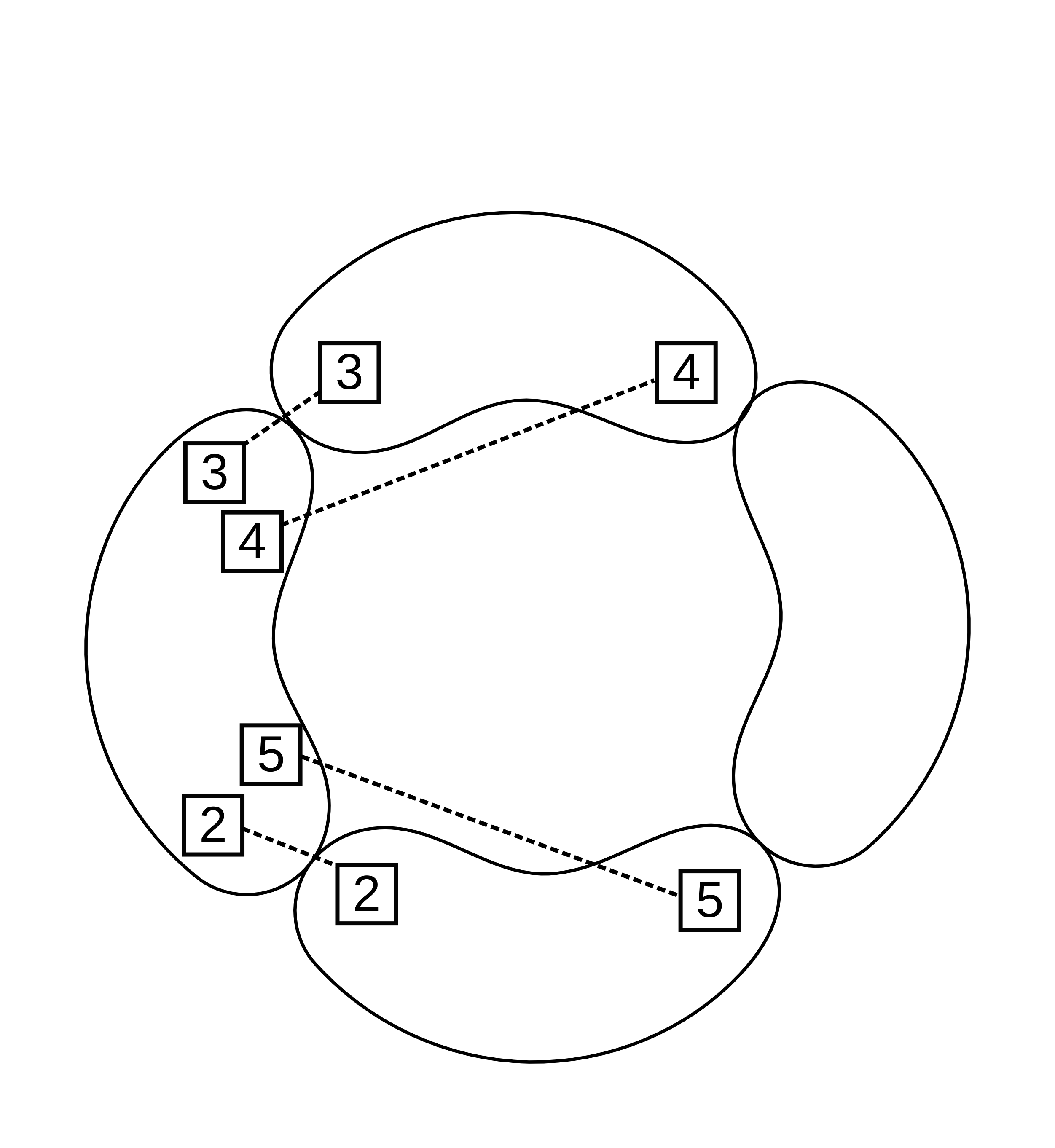}
\begin{tikzpicture}[remember picture,overlay,node distance=4mm, >=latex',block/.style = {draw, rectangle, minimum height=65mm, minimum width=83mm,align=center},]
\node[overlay,align=center] at (2.7,0) {Matter locus 8,\\$L_2, L_3$};
\node[overlay,align=center] at (2.8,4.7) {$\mathbb{P}_{1}^{8,2}$};
\node[overlay,align=center] at (4.55,3.25) {$\mathbb{P}_1^{8,1}$};
\node[overlay,align=center] at (1,3) {$\mathbb{P}_1^{8,4}$};
\node[overlay,align=center] at (2.8,1.35) {$\mathbb{P}_{1}^{8,3}$};
\end{tikzpicture}
\end{minipage}
\vspace{1cm}
\caption{\label{fig:codim2Curve}Splitting of the fiber above loci of codimension two. A number $i$ in a box on a component means that the bisection $[z_i]$ intersects that component.
A double-box is used when the bisection intersects that component twice and the component is shaded when it is wrapped by part of a bisection.}
\end{figure*}
Finally we conclude that the matter spectrum does indeed respect the restriction (\ref{eq:evenweights}) imposed by the torsion Shioda map (\ref{eqn:122_0fullTorsionShioda}).
Only bifundamental matter representations under the non-Abelian gauge group are present.
Moreover the representations $(\one, \two, \two, \one)$ and $(\two,\one,\one,\two)$ both appear over two inequivalent codimension two loci.
From the field theory perspective these are distinguished by the discrete $\mathbb{Z}_4$ symmetry.
The total discrete charge of operators has to be $0\,\text{mod}\,2$ which forbids several Yukawa couplings that would otherwise be allowed.

Two examples of allowed Yukawa couplings are given by the codimension three loci
\begin{align}
\begin{split}
Y_1:&\,L_0\cap L_1\cap\{a_3=0\}\,,\\
Y_2:&\,L_0\cap L_1\cap\{a_4a_7-a_3a_8=0\}\,,
\end{split}
\end{align}
with
\begin{align}
\begin{split}
Y_1:\,&(\two,\two,\one,\one)_{\frac12}\cdot(\two, \one , \two , \one)_{-\frac12}\cdot(\one, \two, \two, \one)_{0} \, ,   \\ 
Y_2:\,&(\two,\two,\one,\one)_{\frac12}\cdot\overline{(\two, \one , \two , \one)_{-\frac12}}\cdot(\one, \two, \two, \one)^\prime_{1} \, .
\end{split}
\end{align}
We find that the $\mathbb{Z}_4$ symmetry not only distinguishes the primed and unprimed representations $(\one, \two, \two, \one)$
but also the two Yukawa couplings that would otherwise be identical due to the pseudo reality of the SU(2) doublet representation.
\section{Outlook}
\label{sec:conclusion}
The observation of a mirror duality between multi-sections and torsion in the sixteen hypersurfaces in two-dimensional toric ambient spaces
could have been blamed on the restricted and fairly symmetric framework.
We have given a combinatorial explanation which crucially depends on special features that are absent in more general fibers.
It is therefore highly non-trivial that the conjecture holds in the far richer class of complete intersections in three-dimensional ambient spaces.
Maybe the two pairs where the degrees of torsion and multi-sections do not match up provide clues to understand this phenomenon.

On the other hand our combinatorial criterion for Mordell-Weil torsion can be used to systematically search for special fibers in complete intersections of higher codimension.
For example it would be interesting to search for models exhibiting $\mathbb{Z}_n\times \mathbb{Z}_m$ Mordell-Weil torsion and check the mirror conjecture in these cases.

Given the number of mirror-pairs in which the conjecture has been verified it
is desirable to pursue a clear mathematical or physical understanding of the
generality of this phenomenon.
\subsection*{Acknowledgments}
We would like to thank Hans Jockers and Albrecht Klemm for useful discussions.
The work of P.O., J.R. and T.S. is partially supported by a scholarship of the Bonn-Cologne Graduate School BCGS, the SFB-Transregio TR33 The Dark Universe (Deutsche Forschungsgemeinschaft) and the European Union 7th network program Unification in the LHC era (PITN-GA-2009-237920). The work of Paul Oehlmann is also supported in part by NSF grant PHY-1417337 and NSF grant PHY-1417316. J.R. would like to thank CERN for hospitality during the completion of this work. P.O. would like to thank KIAS for hospitality and finanical support the completion of this work.

\appendix

\section{Equivalence classes of families}
\label{app:equivalence}
Assume that we have two nef partitions 
\begin{align}
\Delta=\Delta_1^{(1)}+\Delta_2^{(1)}=\Delta_1^{(2)}+\Delta_2^{(2)} \, ,
\end{align}
corresponding to the ambient space $\mathbb{P}_{\Delta^\circ}$ that give identical equations.
In general the two dual nef partitions 
\begin{align}
\nabla^{(1)}=\nabla_1^{(1)}+\nabla_2^{(1)}\quad \text{ and }\quad \nabla^{(2)}={\nabla}_1^{(2)}+{\nabla}_2^{(2)} \, ,
\end{align}
will correspond to different ambient spaces $\mathbb{P}_{{\nabla^{(1)}}^\circ}$ and $\mathbb{P}_{{\nabla^{(2)}}^\circ}$.
To identify the Weierstrass form one can proceed as follows:

For ease of notation we first look at the nef partition $\Delta=\Delta_1^{(1)}+\Delta_2^{(1)},\,\nabla^{(1)}=\nabla_1^{(1)}+\nabla_2^{(1)}$ and drop the superscript.
The Weierstrass form is parametrized by the coefficients of the polynomials $P_{\nabla_1}$ and $P_{\nabla_2}$.
The coefficients correspond to points in the dual lattice $n_i\in \Delta^\circ\cap N$, with $\Delta^\circ=\langle\nabla_1,\,\nabla_2\rangle$.
We will denote the coefficient of the monomial in $P_{\nabla_j}$ that corresponds to the point $0\in\nabla_j$ by $a^j_0$ and take the other coefficients to be $a_k,\,k>0$.
One can use rescalings of the ambient space coordinates to introduce variables that are manifestly invariant under the torus action.

A basis of linear relations $\sum_{i>0} l^s_in_i=0$ among the points $n_i\in(\Delta^\circ\cap N)\backslash 0$ provides a complete set of invariant variables
\begin{align}
z_s=\left(\prod\limits_{i>0}a_i^{l^s_i}\right)/\left[\left(a^1_0\right)^{e^s_1}\left(a^1_0\right)^{e^s_2}\right]\, ,
\label{eqn:invariantvariables}
\end{align}
with
\begin{align}
e^s_j=\sum_{n_i\in(\nabla_j\backslash0)}l^s_i\,.
\end{align}
The linear relations among the points correspond to curves in $\mathbb{P}_{\Delta^\circ}$.
It is well known \cite{Cox:2000vi} that the curves which span the Mori cone of this variety correspond to relations that provide the
large complex structure coordinates for the mirror in $\mathbb{P}_{\nabla^\circ}$.
If there are multiple partitions
\begin{align}
\Delta^\circ=\langle\nabla_1^{(1)},\,\nabla_2^{(1)}\rangle=\langle\nabla_1^{(2)},\,\nabla_2^{(2)}\rangle\, ,
\end{align}
giving multiple mirror manifolds $X^{(1)}\subset\mathbb{P}_{{\nabla^{(1)}}^\circ},\,X^{(2)}\subset\mathbb{P}_{{\nabla^{(2)}}^\circ},$ these coordinates provide
a canonical identification of the complex structure moduli spaces.

Setting $z_s^{(1)}=z_s^{(2)}$ leads to a set of equations
\begin{align}
\frac{A}{B}=\frac{A'}{B'} \, ,
\end{align}
with $A, B$  monomials in the coefficients of $X^{(1)}$ and $A', B'$ monomials in the coefficients of $X^{(2)}$.
Assuming that the sets of coefficients are related by a permutation we can demand $A=B, A'=B'$
and the resulting set of equations is easily solved.
\section{Polynomial non-Abelian loci}
\label{app:nonAbelian}
In the hypersurface case every generic non-Abelian locus of a family arose by setting one of the monomials to zero \cite{Klevers:2014bqa}.
For a section of the anti-canonical bundle of $\mathbb{P}_{\Delta^\circ}$ the possible monomials are in one-to-one correspondence with the points $m\in\Delta\cap M$.
The general hypersurface equation is given by
\begin{align}
P=\sum\limits_{m\in\Delta}\prod\limits_{\rho_i\in\Sigma(1)}s_m z_j^{\langle m,\rho_i\rangle+1} \, .
\end{align}
The polynomial $P$ factorizes after deleting a point $m$ in $\Delta$ if and only if $m$ is the only point which has negative scalar product with a non-zero set of generators
{
\small
\begin{align}
\begin{split}
S_m=\{&\rho\in\Sigma(1)\,:\,\langle m,\,\rho\rangle = -1 \, ,\\
&\langle m',\rho\rangle > -1\text{ for all }m'\in(\Delta\cap M)\backslash m\}\, .
\end{split}
\end{align}
}
This is exactly the case when $m$ is a vertex of $\Delta$ and the dual face in $\Delta^\circ$ has at least one inner point.
The generators in the set $S_m$ are then the inner points of the dual face.
Moreover the associated divisors correspond to the Cartan elements of a factor of the non-Abelian gauge group.
The non-Abelian gauge group can therefore be immediately read off from the polytope.
Faces of $\Delta^\circ$ correspond to $SU(N)$ factors with $N$ being the number of inner points.

For fibers constructed as complete intersections in three-dimensional toric ambient spaces the situation is more complicated and one can in general
not read off the gauge group by looking at the polytope.
Often the singularity of the fiber does not arise from setting one of the coefficients to zero but is given as the vanishing locus of
a more general polynomial $e$.
In many cases these singularities are still torically resolved.
We distinguish three different types of the \textit{polynomial non-Abelian loci}:
\begin{enumerate}
\item In cases where the polynomial $e$ involves coefficients of both complete intersection equations it is in general a product of differences of roots
such that the singularity is again torically resolved.
For example, the equations of \href{http://wwwth.mpp.mpg.de/members/jkeitel/weierstrass/data/8_0.txt}{(8,~0)} are given by
\begin{align}
\begin{split}
p_1 =& a_5z_0^2z_4 + a_4z_0z_3z_4 + a_0z_3^2z_4 + a_3z_1^2z_6\\
&+ a_2z_1z_2z_6 + a_1z_2^2z_6 \, ,\\
p_2 =& a_{11}z_0^2z_4 + a_{10}z_0z_3z_4 + a_6z_3^2z_4 + a_9z_1^2z_6\\
&+ a_8z_1z_2z_6 + a_7z_2^2z_6\, , 
\end{split}
\end{align}
and the classification of \cite{Braun:2014qka} suggests an $I_2$ splitting of the fiber at the codimension one loci
{
\small
\begin{align}
\begin{split}
e_1=a_3^2a_7^2 - a_2a_3a_7a_8 + a_1a_3a_8^2 + a_2^2a_7a_9&\\
-2a_1a_3a_7a_9 - a_1a_2a_8a_9 + a_1^2a_9^2&=0\, ,
\end{split}
\end{align}
}
and
{
\small
\begin{align}
\begin{split}
e_2=a_5^2a_6^2 - a_4a_5a_6a_{10} + a_0a_5a_{10}^2 + a_4^2a_6a_{11}&\\
- 2a_0a_5a_6a_{11} - a_0a_4a_{10}a_{11} + a_0^2a_{11}^2&=0\, .
\end{split}
\end{align}
}
The variables $z_4$ and $z_6$ correspond to rays in the interior of faces of reflexive polytope $8$.
They do not intersect the curve and provide natural candidates for resolving the singularities.
In fact setting $z_4=0$ the two equations become
\begin{align}
\begin{split}
p_1\sim(a_3z_1^2+a_2z_1z_2+a_1z_2^2)\cdot z_6\, ,\\
p_2\sim(a_9z_1^2+a_8z_1z_2+a_7z_2^2)\cdot z_6\, .
\end{split}
\end{align}
The generators corresponding to $z_4$ and $z_6$ do not share a cone but $e_1$ is essentially the product
\begin{align}
e_1\propto\prod\limits_{i,\,j}(r_i-r'_j)\, ,
\end{align}
where $r_i\, ,\,r'_j$ are the roots of $a_3z_1^2+a_2z_1+a_1$ and $a_9z_1^2+a_8z_1+a_7$ respectively.
The exceptional divisor resolving the $I_2$ singularity above the first locus is therefore given by $\{a_3z_1^2+a_2z_1z_2+a_1z_2^2=0\}\cup\{z_4=0\}$.
An analogous relation holds for $e_2$.
\item A second type of splitting occurs when the polynomial only involves coefficients of one of the equations and this equation factors into two polynomials over the non-Abelian locus.
This happens for example in \href{http://wwwth.mpp.mpg.de/members/jkeitel/weierstrass/data/1920_0.txt}{(1920,~0)}, where one of the equations is given by
{
\small
\begin{align}
\begin{split}
p&=a_8z_0z_1+a_5z_4z_5z_6z_7z_8^2z_9^2z_{10}^2z_{11}z_{12}\\
&\phantom{=}+ a_6z_1z_5z_7z_8z_9z_{10}z_{11} + a_7z_0z_4z_6z_8z_9z_{10}z_{12}\, ,
\end{split}
\end{align}
}
and over the locus
\begin{align}
 a_5a_8-a_6a_7=0 \, ,
\end{align}
this factors into
\begin{align}
\begin{split}
p&\sim(a_8 z_1 + a_7 z_{10} z_{12} z_4 z_6 z_8 z_9)\\
&\phantom{\sim} \cdot(a_8 z_0 + a_6 z_{10} z_{11} z_5 z_7 z_8 z_9)\, .
\end{split}
\end{align}
The phenomenon that a K\"ahler deformation on the subvariety is not induced by a toric deformation of the ambient space and therefore appears to be frozen is well
known \cite{Cox:2000vi}. 
We expect that in families where this type of splitting occurs the resolution divisor is non-toric.
There can however be a realization of the same fiber in a different toric ambient space where the resolution is toric.
This is described in Section \ref{sec:redundancy}.
It would be interesting to follow the argument in \cite{Berglund:1998va} to interpret the non-toric resolution not as a deficit in the description but
as a physical freezing of the associated modulus by background flux.
\item Yet another possibility is that the polynomial only involves the coefficients of one of the equations but the equation factors only over a field extension.
The factors are then exchanged under monodromies around higher codimension loci.
An example is given by \href{http://wwwth.mpp.mpg.de/members/jkeitel/weierstrass/data/5_3.txt}{(5,~3)} for which one of the equations reads
\begin{align}
\begin{split}
p =& a_{15}z_2^2 + a_{14}z_2z_3 + a_{12}z_3^2\\
& + a_{13}z_2z_4+ a_{11}z_3z_4 + a_{10}z_4^2 \, .
\end{split}
\end{align}
At the locus
\begin{align}
\begin{split}
-&a_{12}a_{13}^2 + a_{11}a_{13}a_{14} - a_{10}a_{14}^2\\
- &a_{11}^2a_{15} + 4a_{10}a_{12}a_{15}=0 \, ,
\end{split}
\end{align}
this splits into
\begin{align}
\begin{split}
p&\sim(a_{15}z_2+b_+z_3+c_+z_4)\\
&\phantom{\sim}\cdot(a_{15}z_2+b_-z_3+c_-z_4)\, ,
\end{split}
\end{align}
with
\begin{align}
\begin{split}
b_\pm=&\frac{b_0\pm b_1\sqrt{d}}{2(a_{13}^2-4a_{10}a_{15})}\, ,\\
c_\pm=&\frac{1}{2}\left(a_{13}\pm\sqrt{d}\right)\, ,
\end{split}
\end{align}
and
\begin{align}
\begin{split}
b_0=&a_{13}^2a_{14}-4a_{10}a_{14}a_{15}\, ,\\
b_1=&a_{13}a_{14}-2a_{11}a_{15}\, ,\\
d=&a_{13}^2-4a_{10}a_{15}\, .
\end{split}
\end{align}
It can be seen that the two components of what appears to be an $I_2$ fiber are exchanged under monodromies
around the locus $d=0$ .
\end{enumerate}
\bibliographystyle{utphys}	
\bibliography{ref}

\end{document}